\documentclass[prd,aps,floatfix,nofootinbib,twocolumn,tightenlines,showpacs]{revtex4}
\usepackage{dcolumn}% Align table columns on decimal point
\usepackage{amsmath}
\usepackage{latexsym}
\usepackage{graphicx}
\usepackage{bm}
\def\co{{\cal O}}

\def\tbeta{\tilde\beta}

\def\bx{{\bf x}}

\def\svev#1{\left\langle #1\right\rangle}       % variable < >
\def\tr{{\rm tr}\,}

\long \def \blockcomment #1\endcomment{}

\def\Eq#1{Eq.~(\ref{#1})}

\newcommand{\bee}{\begin{equation}}
\newcommand{\ee}{\end{equation}}
\newcommand{\beea}{\begin{eqnarray}}
\newcommand{\eea}{\end{eqnarray}}

\begin{document}
%%%%%%%%%%%%%%%%%%%%%%%%%%%%%%%%%%%%%%%%%%%%%%%%%%%%%%%%%%%%%%%%%%%%%%
\title{
Infrared fixed point in SU(2) gauge theory with adjoint fermions
}
\author{Thomas DeGrand}%
%\email{degrand@pizero.colorado.edu}
 \affiliation{Department of Physics,
University of Colorado, Boulder, CO 80309, USA}
\author{Yigal Shamir}
%\email{shamir@post.tau.ac.il}
\author{Benjamin Svetitsky}
%\email{bqs@julian.tau.ac.il}
\affiliation{Raymond and Beverly Sackler School of Physics and Astronomy,
Tel~Aviv University, 69978
Tel~Aviv, Israel}

\begin{abstract}
We apply Schr\"odinger-functional techniques to the SU(2) lattice gauge theory
with $N_f=2$ flavors of fermions in the adjoint representation.
Our use of hypercubic smearing enables us to work at stronger couplings
than did previous studies, before encountering a critical point and a bulk
phase boundary.  Measurement of the running coupling constant
gives evidence of an infrared fixed point $g_*$ where $1/g^2_*=0.20(4)(3)$.
At the fixed point, we find a mass anomalous dimension
$\gamma_m(g_*)=0.31(6)$.
\end{abstract}

\maketitle

%%%%%%%%%%%%%%%%%%%%%%%%%%%%%%%%%%%%%%%%%%%%%%%%%%%%%%%%%%%%%%%%%%%%%
\section{Introduction}
%%%%%%%%%%%%%%%%%%%%%%%%%%%%%%%%%%%%%%%%%%%%%%%%%%%%%%%%%%%%%%%%%%%%%

Recently, there has been an upsurge in interest in applying the techniques of
lattice gauge theory
to theories with different numbers of colors, different numbers of flavors, and fermionic
 representations other than the fundamental \cite{Fleming:2008gy,Pallante:2009hu,Sardinia,DeGrand:2010ba,Rummukainen:2011xv}.
The immediate phenomenological application of these studies is to candidate theories
of physics beyond the Standard Model, in which new gauge dynamics  allow the
replacement of  the fundamental Higgs boson by a bound state of new fermionic degrees of freedom---so-called ``technicolor'' theories \cite{Hill:2002ap}.
More generally, it is an interesting question in quantum field theory: How do
systems of gauge fields coupled to fermions behave?

Many studies attempt to characterize the behavior of a theory by computing
a suitably defined running coupling constant.
Perturbation theory \cite{Caswell:1974gg,Banks:1981nn} gives us a first approach
and a menu of possibilities to be confronted by
numerical studies. The two-loop beta function is
\bee
\beta(g^2)=\frac{dg^2}{d\log \mu^2}=-\frac{b_1}{16\pi^2}g^4-\frac{b_2}{(16\pi^2)^2}g^6,
\label{eq:2loopbeta}
\ee
where, for an SU($N$) gauge theory with $N_f$ flavors of fermions in representation $R$,
\beea
b_1&=&  \frac{11}{3}\, C_2(G) - \frac{4}{3}\,N_f T(R), \label{2loopbeta1}\\
b_2&=& \frac{34}{3} [C_2(G)]^2
  -N_f T(R) \left[\frac{20}{3} C_2(G) %\right.\nonumber\\&&
  + 4 C_2(R) \right].
\label{eq:2loopbeta2}
\eea
Here $C_2(R)$ is the value of the quadratic Casimir operator in representation $R$ [where
$G$ denotes the adjoint representation, so $C_2(G)=N$], while $T(R)$ is the conventional trace normalization.
Three possibilities for the behavior of the massless theory are:
\begin{itemize}
\item triviality, $b_1<0$ in our conventions.
\item QCD-like physics, $\beta(g^2)<0$ for all $g^2$, meaning $b_1,b_2>0$ in \Eq{eq:2loopbeta}.
This  is presumed to be associated with
confinement and chiral symmetry breaking, as in ordinary QCD.
\item a fixed point at $g=g_*$, such that $\beta(g_*^2)=0$. For $b_1>0$ this comes about if $b_2<0$ and then $g_*$ is an infrared-attractive
fixed point (IRFP). Correlation functions decay algebraically at large distance,
and there is  no  confinement,
no chiral symmetry breaking, and indeed no particle spectrum.
For a given representation $R$,
the domain of $(N_c,N_f)$  where an IRFP exists is called the ``conformal window.''
\end{itemize}
Technicolor demands a theory with QCD-like physics.
{\em Walking\/} technicolor requires a QCD-like theory just outside the conformal window.

In a massive theory, the running coupling $g(\mu)$ is augmented by the running fermion mass $m(\mu)$.
In addition to the beta function, one considers the anomalous dimension $\gamma_m$ of the mass operator $\bar\psi\psi$.
It determines the running of the mass parameter according to
\bee
\mu \frac{dm(\mu)}{d\mu} = -\gamma_m(g^2) m(\mu).
\ee
In lowest order in perturbation theory,
\bee
\gamma_m= \frac{6 C_2(R)}{16\pi^2} g^2.
\ee
In the massless theories used for technicolor, $\gamma_m$ governs the
 running of the condensate $\svev{\bar\psi\psi}$.
It is thus an important diagnostic for realistic ``extended'' technicolor models, especially those based on walking.
Phenomenological constraints on such models require it to have a large, nonperturbative value.
If the massless theory is conformal, on the other hand, then near $m_q=0$ the correlation length $\xi$ scales as
\bee
\xi \sim m_q^{-{1}/{y_m}}
\ee
where $y_m = 1+\gamma_m(g_*)$ is the leading relevant exponent of
 the system~\cite{DeGrand:2009hu,DelDebbio:2010ze}.
(Here $m_q$ is the mass defined by the axial Ward identity---see below.)

Perturbation theory cannot make reliable predictions for properties of systems that
evolve to become strongly interacting at long distances.
To address questions such as ``where is the conformal window'' and ``what is the spectrum'' (for a confining theory) or ``what are the critical exponents'' (for a conformal theory) requires nonperturbative methods.
For us, this methodology is simulation of the lattice-regulated theory.

We present here a study of the SU(2) gauge theory coupled to two flavors of
Dirac fermions in the adjoint representation of the gauge group.
Following the introduction of this theory as a technicolor candidate \cite{Sannino:2004qp,Dietrich:2006cm}, several groups have performed numerical simulations
\cite{Catterall:2007yx,DelDebbio:2008zf,Catterall:2008qk,Hietanen:2008mr,Hietanen:2009az,DelDebbio:2009fd,Bursa:2009we,Catterall:2009sb,Catterall:2010du,DelDebbio:2010hu,DelDebbio:2010hx}.
Studies of the lattice theory's phase diagram \cite{Catterall:2008qk,Hietanen:2008mr}
and spectrum \cite{Catterall:2007yx,DelDebbio:2008zf,Catterall:2008qk,Hietanen:2008mr,DelDebbio:2009fd,Catterall:2009sb,DelDebbio:2010hu}
indicate that its weak-coupling phase is quite different from that of
SU(2) or SU(3) gauge theories coupled
to a small number of fundamental-representation fermions.
Applications of the Schr\"odinger functional (SF) method \cite{Hietanen:2009az,Bursa:2009we} and Monte Carlo renormalization group (MCRG)
\cite{Catterall:2010du} have
indicated that this theory has  an IRFP.
The reliability of this prediction is limited, however, by the lattice discretization used.
The phase diagram obtained shows a first-order transition in strong coupling which terminates 
rather close to the claimed location of the IRFP.
One consequence is that the value of $g_*$, if it exists, is poorly determined.

The main goal of our work was to determine whether this theory has an infrared fixed point.
We have calculated the beta function for the SF coupling $g^2$ by carrying out simulations
 on a number of different lattice volumes.
We are able to present strong, even definite evidence for an IRFP.
The strength of our calculation comes from adoption of an improved fermion action that
 incorporates normalized hypercubic smearing, ``nHYP fat links'' \cite{Hasenfratz:2001hp,Hasenfratz:2007rf}.
The fat-link action has been shown to effect a dramatic reduction of lattice artifacts when
 used for QCD simulations.
As we found for the SU(3) theory with sextet fermions \cite{DeGrand:2010na}, this action 
 moves the end point of the first order transition farther into strong coupling and allows
 us to examine a much larger range of the SF coupling without encountering it.
The work of Hietanen {\em et al.}~\cite{Hietanen:2009az} was limited to the range $1/g^2> 0.22$
while the estimate for $g_*^2$ was in the range 2.0 to 3.2 or $1/g_*^2=0.3$ to 0.5.
We reach $1/g^2\simeq0.07$ to 0.10, depending on the volume, at the strongest
bare coupling.
(Here the beta function has changed sign, so the largest
running coupling is obtained on the smallest volume.)
We observe directly an IRFP at $1/g_*^2=0.20(4)(3)$ where
the first error is statistical and the second is systematic.
This is a slightly weaker coupling than the two-loop perturbative
 value of  $g_*^2=7.9$ or $1/g_*^2=0.126$.
With this result, we determine the value of the mass anomalous
dimension at the IRFP, $\gamma_m(g_*)=0.31(6)$; here the bulk
of the error reflects the uncertainty in $g_*$. 

The outline of the paper is as follows: In Sec.~\ref{sec:method} we review our lattice action
and the techniques we use to measure the beta function and $\gamma_m$.
In Sec.~\ref{sec:phases} we present the phase diagram of the lattice theory.
Sections~\ref{sec:SF} and~\ref{sec:zp} contain our results for the running coupling constant
 and mass anomalous dimension.
We discuss our results in the context of the literature in Sec.~\ref{sec:last}.

%%%%%%%%%%%%%%%%%%%%%%%%%%%%%%%%%%%%%%%%%%%%%%%%%%%%%%%%%%%%%%%%%%%%%
\section{Method \label{sec:method}}
%%%%%%%%%%%%%%%%%%%%%%%%%%%%%%%%%%%%%%%%%%%%%%%%%%%%%%%%%%%%%%%%%%%%%
The lattice calculations in our SU(2) study, following the Schr\"odinger functional method, were carried out in the same way as in our recent study
of the SU(3) gauge theory with sextet fermions.
We refer the reader to Ref.~\cite{DeGrand:2010na} for a detailed presentation.
In Secs.~\ref{ssec:action} and~\ref{ssec:SF}  we give a short recapitulation to make this paper self-contained.
In Sec.~\ref{ssec:extraction} we describe at some length the extraction of the beta function from fits to the running coupling.
The method used here is special to a theory with a slowly running coupling.
Our method for calculating $\gamma_m$, again similar to that in our SU(3) work, is presented in Sec.~\ref{ssec:gamma}.

%%%%%%%%%%%%%%%%%%%%%%%%%%%%%%%%%%%%%%%%%%%%%%%%%%%%%%%%%%%%%%%%%%%%%
\subsection{Lattice action and simulation \label{ssec:action}}
%%%%%%%%%%%%%%%%%%%%%%%%%%%%%%%%%%%%%%%%%%%%%%%%%%%%%%%%%%%%%%%%%%%%%

We study the SU(2) gauge theory coupled to two flavors of dynamical fermions in the
adjoint representation of the color gauge group.
The lattice action is given by the single-plaquette gauge action and the
Wilson fermion action with added clover term~\cite{Sheikholeslami:1985ij}.
The gauge connections in the fermion action employ the differentiable hypercubic
smeared link of Ref.~\cite{Hasenfratz:2001hp}, from which the adjoint-representation
gauge connection for the fermion operator is constructed.
[The adaptation of the original SU(3) construction to SU(2) is trivial.]
The smearing parameters for the links are the same as in
Ref.~\cite{Hasenfratz:2001hp}: $\alpha_1=0.75$, $\alpha_2=0.6$, $\alpha_3=0.3$.
The parameters that are inputs to the simulation are the bare gauge coupling $\beta=4/g_0^2$
and the fermion hopping parameter $\kappa$, related to the bare mass $m_0$ by
$\kappa=(8+2m_0)^{-1}$.
Tests of nonperturbative improvement \cite{Capitani:2006ni,Durr:2008rw,Shamir:2010cq} indicate that
we can safely set the clover coefficient to its tree-level value of unity.

The molecular dynamics integration is accelerated with an additional heavy pseudo-fermion
field as suggested by Hasenbusch~\cite{Hasenbusch:2001ne}, multiple time scales~\cite{Urbach:2005ji},
and a second-order Omelyan integrator~\cite{Takaishi:2005tz}.
Lattice sizes range from $6^4$ to $16^4$ sites.

We study the massless theory by fixing $\kappa=\kappa_c(\beta)$,
the point at which the quark mass $m_q$
 vanishes for each $\beta$. We define $m_q$ using the unimproved axial Ward identity (AWI),
\bee
\partial_t \sum_\bx \svev{A_0^a(\bx,t)\co^a} = 2m_q \sum_\bx \svev{ P^a(\bx,t)\co^a}.
\label{eq:AWI}
\ee
where the axial current $A_\mu^a=\bar \psi \gamma_\mu\gamma_5 (\tau^a/2)\psi$, the pseudoscalar density $P^a=\bar \psi \gamma_5 (\tau^a/2)\psi$, and $\co^a$ could
be any source. We follow the usual SF procedure and take the source to be the gauge-invariant
wall source at $t=a$ as in \Eq{eq:fPdef} below.
The correlation functions in \Eq{eq:AWI} are then measured at $t=L/2$, the midpoint of the
 lattice.
The derivative is taken as the symmetric difference, $\partial_\mu f(x)=[f(x+\hat\mu a) -
f(x-\hat\mu a)]/(2a)$.

On a finite lattice, the quark mass $m_q$ generally depends on the lattice size
$L$ as well as on $(\beta,\kappa)$.
As we show below, the dependence on $L$ is quite weak except at the strongest
couplings.
We generally defined $\kappa_c$ by demanding $m_q=0$ on a $12^4$ lattice.
Since the $L$ dependence becomes significant at strong coupling (see the Appendix),
we also carried out a complete SF calculation at a shifted
$\kappa$ for $\beta=1.4$.
As will be seen below, the results turned out to be insensitive to this shift.

%%%%%%%%%%%%%%%%%%%%%%%%%%%%%%%%%%%%%%%%%%%%%%%%%%%%%%%%%%%%%%%%%%%%%
\subsection{Schr\"odinger functional and the running coupling \label{ssec:SF}}
%%%%%%%%%%%%%%%%%%%%%%%%%%%%%%%%%%%%%%%%%%%%%%%%%%%%%%%%%%%%%%%%%%%%%
The Schr\"odinger functional (SF)~\cite{Luscher:1992an,Luscher:1993gh,Sint:1995ch,Jansen:1998mx,DellaMorte:2004bc}
 is an implementation of the background field method that is especially suited for  lattice
 calculations. It involves performing simulations in a finite volume of linear
dimension $L$, while imposing fixed boundary conditions on the gauge field.
The classical field that minimizes the Yang--Mills action
subject to these boundary conditions is a background color-electric field.
By construction the only distance scale that characterizes the background field is $L$, so
the $n$-loop effective action $\Gamma\equiv-\log Z$ gives the running coupling via
\bee
\label{Gamma}
\Gamma = g(L)^{-2} S_{\textrm{YM}}^{\textrm{cl}} ,
\ee
where
$S_{\textrm{YM}}^{\textrm{cl}}$
is the classical action of the background field.
When $\Gamma$ is calculated non-perturbatively, \Eq{Gamma} gives a non-perturbative
definition of the running coupling at scale $L$.
In a simulation, the coupling constant is determined through the
differentiation of \Eq{Gamma} with respect to some parameter $\eta$ in the boundary conditions.
This is an observable quantity,
\begin{eqnarray}
  \left.\frac{\partial \Gamma}{\partial\eta} \right|_{\eta=0}
  &=&
  \left.\svev{\frac{\partial S_{YM}}{\partial\eta}
  -\tr \left( \frac{1}{D_F^\dagger}\;
        \frac{\partial (D_F^\dagger D_F)}{\partial\eta}\;
            \frac{1}{D_F} \right)}\right|_{\eta=0} \nonumber\\
&=& \frac{K}{g^2(L)} .
            \label{deta}
\end{eqnarray}
The quantity $K$ is just a number \cite{Luscher:1993gh}.
With boundary fields as described in Ref.~\cite{Jansen:1998mx}, it takes the value $K=-12\pi$ in the infinite-volume limit; we use this value for finite volume as well, since the corrections are numerically small.
We also impose twisted spatial boundary conditions on the fermion fields, following
 Ref.~\cite{Sint:1995ch},
$\psi(x+L)=\exp(i\theta)\psi(x)$, with $\theta=\pi/5$ on all three
axes \cite{DellaMorte:2004bc}.

The observable (\ref{deta}) is quite noisy and requires long simulation runs, as shown in Table~\ref{tab:runs}.

%%%%%%%%%%%%%%%%%%%%%%%%%%%%%%%%%%%%%%%%%%%%%%%%%%%%%%%%%%%%%%%%%%%%%
\begin{table}
\caption{Summary of simulation runs for obtaining the
Schr\"odinger functional coupling $g^2$ at the bare couplings
 $(\beta, \kappa_c)$, for the lattice sizes $L$ used in this study.
Trajectories were of unit length.}
\begin{center}
\begin{ruledtabular}
\begin{tabular}{ddrrrr}
\beta & \kappa_c &\multicolumn{4}{c}{trajectories}\\
\cline{3-6}
&&              $L=6a$    & $L=8a$    & $L=12a$   & $L=16a$  \\
\hline
3.0 & 0.12682 &  16.2K &  32.2K & 30K & 19.5K  \\
2.5 & 0.1276 &  16.2K & 32.2K  &  40.6K & 27.2K  \\
2.453 & 0.12766 & 16.2K  &  \hfil-- &  16.2K &   \hfil--  \\
2.445 & 0.12769 & 16.2K  &  16.2K & \hfil-- &   \hfil--  \\
2.0 & 0.12967 & 16.2K & 32.2K & 41.6K & 27.2K  \\
1.985 & 0.12279 & 16.2K  &  \hfil-- & 16.2K  &   \hfil--  \\
1.97 & 0.12991 & 16.2K & 16.2K & \hfil-- &   \hfil--  \\
1.75 & 0.13216 & 16.2K & 32.2K & 32.3K &  41.8K \\
1.5 & 0.13617 & 16.2K  & 32.2K  & 46.2K & 32.3K \\
1.4 & 0.13824 & 16.2K & 32.2K & 43K & 33.2K \\
\end{tabular}
\end{ruledtabular}
\end{center}
\label{tab:runs}
\end{table}
%%%%%%%%%%%%%%%%%%%%%%%%%%%%%%%%%%%%%%%%%%%%%%%%%%%%%%%%%%%%%%%%%%%%%

%%%%%%%%%%%%%%%%%%%%%%%%%%%%%%%%%%%%%%%%%%%%%%%%%%%%%%%%%%%%%%%%%%%%%
\subsection{Extraction of the beta function \label{ssec:extraction}}
%%%%%%%%%%%%%%%%%%%%%%%%%%%%%%%%%%%%%%%%%%%%%%%%%%%%%%%%%%%%%%%%%%%%%
By calculating the inverse running coupling
on lattices of size $L$ and $sL$, we obtain the discrete beta function (DBF)
\bee
B(u,s) = \frac1{g^2(sL)}-\frac1{g^2(L)}, \qquad u \equiv \frac1{g^2(L)}\ .
\label{DBF}
\ee
The usual beta function refers to infinitesimal scale changes.
We define the beta function for the inverse coupling as
\bee
  \tbeta(1/g^2) \equiv \frac{d(1/g^2)}{d\log L}
  = 2\beta(g^2)/g^4 = 2u^2 \beta(1/u).
\label{invbeta}
\ee
Hence,
\bee
\log s = \int_{L}^{sL} \frac{dL'}{L'} =
\int_{u}^{u+B(u,s)}\frac{du'}{\tbeta(u')}\ .
\label{DBF2}
\ee

While the literature is careful to distinguish between the DBF
and the usual beta function,
we remark that in our theory the DBF's we can measure are, to high accuracy,
just proportional to the beta function itself. This occurs for two reasons.
First, our coupling runs slowly: We are, after all, near the (anticipated)
bottom of the conformal window.
Second, due to the cost of simulations,
monitoring the volume dependence of the running coupling at fixed
bare parameters is practical only for $s\alt2$.

\begin{figure}
\begin{center}
\includegraphics[width=\columnwidth,clip]{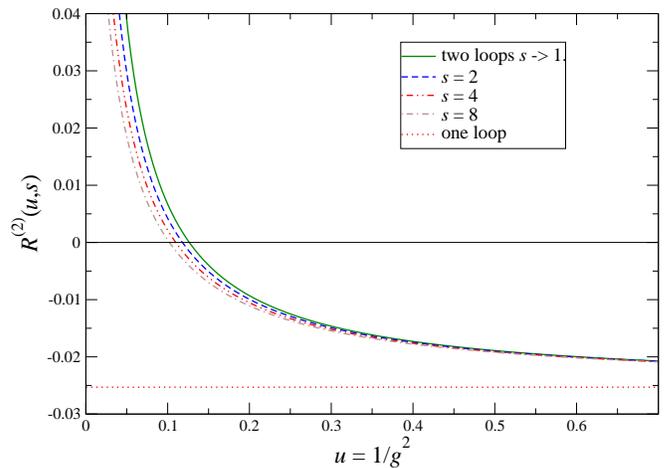}
\end{center}
\caption{Rescaled discrete beta function,
calculated in two loops for various scaling factors $s$.
Also shown are the one- and two-loop beta functions;
the rescaled DBF approaches the two-loop beta function when $s\to1$.
Top to bottom, the curves are in the order shown in the legend.
\label{fig:bfn}}
\end{figure}

If the beta function changes little in the course of
integrating Eq.~(\ref{DBF2}),
then the \textit{rescaled} DBF, defined as
\bee
R(u,s)=\frac{B(u,s)}{\log s}\ ,
\label{RDBF}
\ee
will be approximately equal to the beta function $\tbeta(u)$.
At one-loop order $R^{(1)}(u,s)=-2b_1/16\pi^2$, a constant
[compare Eqs.~(\ref{eq:2loopbeta}) and~(\ref{invbeta})].
The situation at the next order is illustrated in Fig.~\ref{fig:bfn}.
The figure shows the two-loop result,
\begin{eqnarray}
R^{(2)}(u,s) &=& -\frac{2b_1}{16\pi^2}- \frac{b_2}{16\pi^2b_1} 
\nonumber\\
&&\times\frac{\log\left[1+(2b_1/16
\pi^2) u^{-1} \log s\right]}{\log s}\ ,
\label{2loopggen}
\end{eqnarray}
for the rescaled DBF for scale factor $s=2$, 4, 8, compared to the
one-loop and two-loop beta functions.%
\footnote{In this model, $b_1=2$ and $b_2=-40$.}
The rescaled DBF for $s=2$ is hardly distinguishable from the beta function.

There are two lessons to be drawn from Fig.~\ref{fig:bfn}.
If the actual DBF resembles the two-loop result,
we can combine the rescaled DBF's for many scale factors $s$ onto a single plot
to give a good approximation to the actual beta function.
Furthermore, since any value of $s\alt2$ is as good as another,
we can combine the couplings for all lattice volumes studied to extract the
beta function via a fit.

We do not have to rely on perturbation theory when going beyond
the approximation of a constant rescaled DBF\@.  Expand the beta function
$\tbeta(u)$ linearly near some fiducial value $u_1$,
\bee
\frac{du}{d\log s} = \tilde\beta(u_1) + (u-u_1)\tilde\beta'(u_1) =B_0 + B_1 u.
\label{approxbeta}
\ee
Upon integration,
\bee
u(s)-u_1 = \tilde\beta(u_1)\frac{\exp(B_1 \log s) -1}{B_1}\ .
\label{linearization}
\ee
When the product $B_1 \log s$ is small the exponential can be expanded,
and we again observe that $[u(s)-u_1]/\log s=R(u,s)$ gives the beta function.

%%%%%%%%%%%%%%%%%%%%%%%%%%%%%%%%%%%%%%%%%%%%%%%%%%%%%%%%%%%%%%%%%%%%%
%%%%%%%%%%%%%%%%%%%%%%%%%%%%%%%%%%%%%%%%%%%%%%%%%%%%%%%%%%%%%%%%%%%%%

This discussion suggests that we can fit the running coupling
from all volumes at fixed bare coupling to
\bee
1/g^2(sL_0) = c_0 +c_1\log s +c_2(\log s)^2 +\cdots,
\label{eq:fit1}
\ee
where $L_0$ is a fixed reference volume.
We treat all the parameters $c_0,c_1,c_2,\cdots$, as independent,
having in mind that Eq.~(\ref{approxbeta}) is in itself just an approximation.
If the terms nonlinear in $\log s$ are small, the slope $c_1$ gives
the reduced DBF directly, according to \Eq{RDBF}. We do this below.
The success of this analysis shows again that the beta function in this
theory is small in the region studied;
we use other fits to estimate the systematic error.

The discussion so far has ignored discretization errors.
The usual analysis found in the literature
presents the beta function only after an extrapolation to $(a/L)\rightarrow 0$.
This is done by collecting data for DBF's on different volumes,
and at the same value of $g^2$. We can do this by performing simulations
at one value of the bare coupling $\beta$
on our largest lattices ($16^4$ volumes),
then moving to smaller volumes, matching
$1/g^2(L)$ at fiducial volumes and measuring $1/g^2(sL)$
on appropriate larger volumes.
For example, with $s=2$ we can look at $L=16$ and 8 at $\beta=2.5$
(for example) compared to $L=12$ and 6 at slightly offset $\beta$ values.
We must move along the $\kappa_c$ line as we do this.
We have attempted this at two values of the SF coupling.  We found (see below)
that the shifts in $\beta$ were not well determined with
the statistics available.
This is, in fact, a disadvantage of having such a small beta function.

Such an analysis, however, is needlessly complicated for the question
we set out to answer, namely, is there an IRFP, and, if so, where is it?
We can test various hypotheses for the dependence of
the running coupling on the lattice
volume at any fixed bare coupling.  Equation (\ref{eq:fit1})
assumes that this dependence reflects continuum physics only; in addition,
we will try fit functions that test
if the volume dependence can be explained by the anticipated form of
discretization errors, \textit{i.e.}, powers of $a/L$.
We can also test for the presence of lattice artifacts by
varying the data sets kept in the fit.
If the results agree on the existence and
location of a zero of the DBF, then we can claim to have found
an IRFP.

%%%%%%%%%%%%%%%%%%%%%%%%%%%%%%%%%%%%%%%%%%%%%%%%%%%%%%%%%%%%%%%%%%%%%
\subsection{Mass anomalous dimension \label{ssec:gamma}}
%%%%%%%%%%%%%%%%%%%%%%%%%%%%%%%%%%%%%%%%%%%%%%%%%%%%%%%%%%%%%%%%%%%%%

The volume dependence of the renormalization factor $Z_P$ of the
isovector pseudoscalar density $P^a=\bar\psi\gamma_5(\tau^a/2)\psi$ gives
the mass anomalous dimension $\gamma_m$.
(The pseudoscalar density is related by a chiral rotation to $\bar\psi\psi$,
which is the object of interest.) It is computed from two correlators
via~\cite{Sint:1998iq,Capitani:1998mq,DellaMorte:2005kg,Bursa:2009we}
\bee
Z_P = \frac {c \sqrt{f_1}}{f_P(L/2)}.
\label{eq:ZP}
\ee
$f_P$ is the propagator from the $t=0$ boundary to a point pseudoscalar
operator at time $x_0$,
\beea
  f_P(x_0)&=&-\frac{1}{3}\sum_a \int d^3y\, d^3z\,  \left\langle
  \overline{\psi}(x_0)\gamma_5\frac{\tau^a}{2}\psi(x_0)\right.\nonumber\\
  &&\times\left.\overline{\zeta}(y)\gamma_5\frac{\tau^a}{2}\zeta(z)
  \right\rangle.
  \label{eq:fPdef}
\eea
We take $x_0=L/2$. In the expression,  $\zeta$ and $\bar\zeta$ are
gauge-invariant wall sources at $t=a$, i.~e., one lattice layer away from
the $t=0$ boundary.  The $f_1$ factor is the boundary-to-boundary correlator,
which cancels the normalization of the wall source. Explicitly, it is
\beea
  f_1&=&-\frac{1}{3L^6}\sum_a \int d^3u\, d^3v\, d^3y\, d^3z\,
  \left\langle
  \overline{\zeta}^\prime(u)\gamma_5 \frac{\tau^a}{2}{\zeta}^\prime(v)
  \right.\nonumber\\
  &&\times\left.\overline{\zeta}(y)\gamma_5\frac{\tau^a}{2}\zeta(z)
  \right\rangle,
  \label{eq:f1def}
\eea
where $\zeta'$ and $\bar\zeta'$ are wall sources at $t=L-a$.

We use the same boundary conditions for the calculation of $Z_P$
as for the simulations that generate the data for the SF coupling.
This makes its computation parasitic on the SF runs.

The constant $c$ allows imposing a volume-independent normalization condition
in the weak-coupling limit. Since we will only need ratios of values of $Z_P$
to find $\gamma_m$, the overall normalization is irrelevant.
We set $c=1/\sqrt2$ in tabulating $Z_P$ below.

We extract the anomalous dimension of $\bar\psi\psi$
from the change in $Z_P$ [\Eq{eq:ZP}]
 between systems rescaled as $L\to sL$.
The (continuum) mass step scaling
function~\cite{Sint:1998iq,Capitani:1998mq,DellaMorte:2005kg,Bursa:2009we} is
\bee
  \label{eq:sigma_p}
  \sigma_P(v,s) = \left. {\frac{Z_P(sL)}{Z_P(L)}}
  \right|_{g^2(L)=v}.
\ee
It is related to the mass anomalous dimension via
\bee
\label{eq:sigPgamma}
  \sigma_P(v,s) = \exp\left[-\int_1^s \frac {dt}{t}\,
  \gamma_m\left(g^2(tL)\right)\right] .
\ee
Because the SF coupling $g^2(L)$ runs so slowly,
\Eq{eq:sigPgamma} is well approximated by
\bee
\sigma_P(g^2,s) = s^{ - \gamma_m(g^2)}.
  \label{eq:gamma}
\ee
We can therefore combine many $sL$ values collected at the same bare parameter
values into one fit function giving $\gamma_m$,
\bee
\log Z_P(L)=-\gamma_m \log L +\text{const}.
  \label{eq:lgamma}
\ee
This fitting procedure parallels keeping only the $c_0$ and $c_1$
terms in \Eq{eq:fit1}.  As in the case of the DBF, we can look for
subleading continuum corrections and/or for lattice artifacts
by modifying the fit functions or the data set kept in the fit.

%%%%%%%%%%%%%%%%%%%%%%%%%%%%%%%%%%%%%%%%%%%%%%%%%%%%%%%%%%%%%%%%%%%%%
\begin{figure}
\begin{center}
\includegraphics[width=\columnwidth,clip]{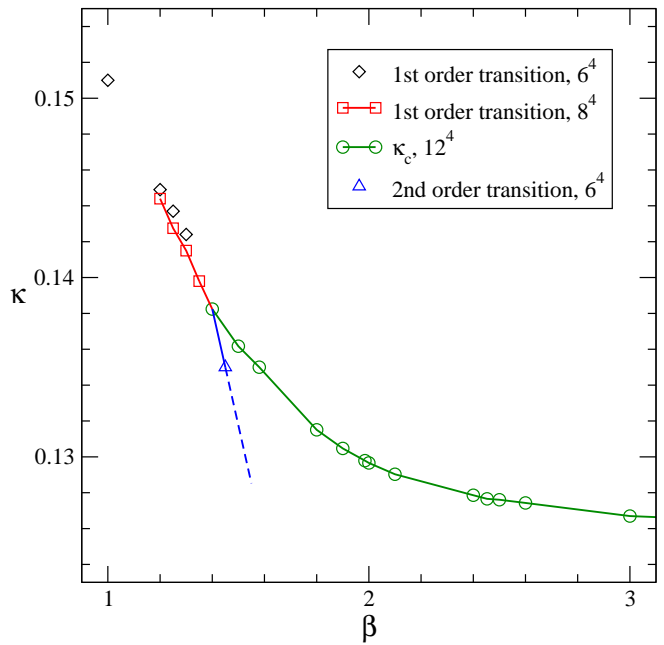}
\end{center}
\caption{Phase diagram in the $\beta$--$\kappa$ plane, determined with
Schr\"odinger functional boundary conditions.
The $\kappa_c$ line ends at a point $(\beta^*,\kappa^*)$
presumed to be a critical point, where
the indicated first-order boundary ends as well.
The triangle marks a point on the second-order phase boundary between
confined and deconfined phases. 
This phase boundary is presumed to continue all the way to $\kappa=0$.
Its upper endpoint may coincide with $(\beta^*,\kappa^*)$.
\label{fig:bk}}
\end{figure}
%%%%%%%%%%%%%%%%%%%%%%%%%%%%%%%%%%%%%%%%%%%%%%%%%%%%%%%%%%%%%%%%%%%%%

%%%%%%%%%%%%%%%%%%%%%%%%%%%%%%%%%%%%%%%%%%%%%%%%%%%%%%%%%%%%%%%%%%%%%
\section{Phase diagram \label{sec:phases}}
%%%%%%%%%%%%%%%%%%%%%%%%%%%%%%%%%%%%%%%%%%%%%%%%%%%%%%%%%%%%%%%%%%%%%

Our determination of the phase diagram in the $(\beta,\kappa)$ plane is preliminary.
The diagram is qualitatively consistent with that given in Refs.~\cite{Catterall:2008qk,Hietanen:2008mr}, but of course it is quantitatively different because of the fat-link action.

%%%%%%%%%%%%%%%%%%%%%%%%%%%%%%%%%%%%%%%%%%%%%%%%%%%%%%%%%%%%%%%%%%%%%
\begin{figure*}
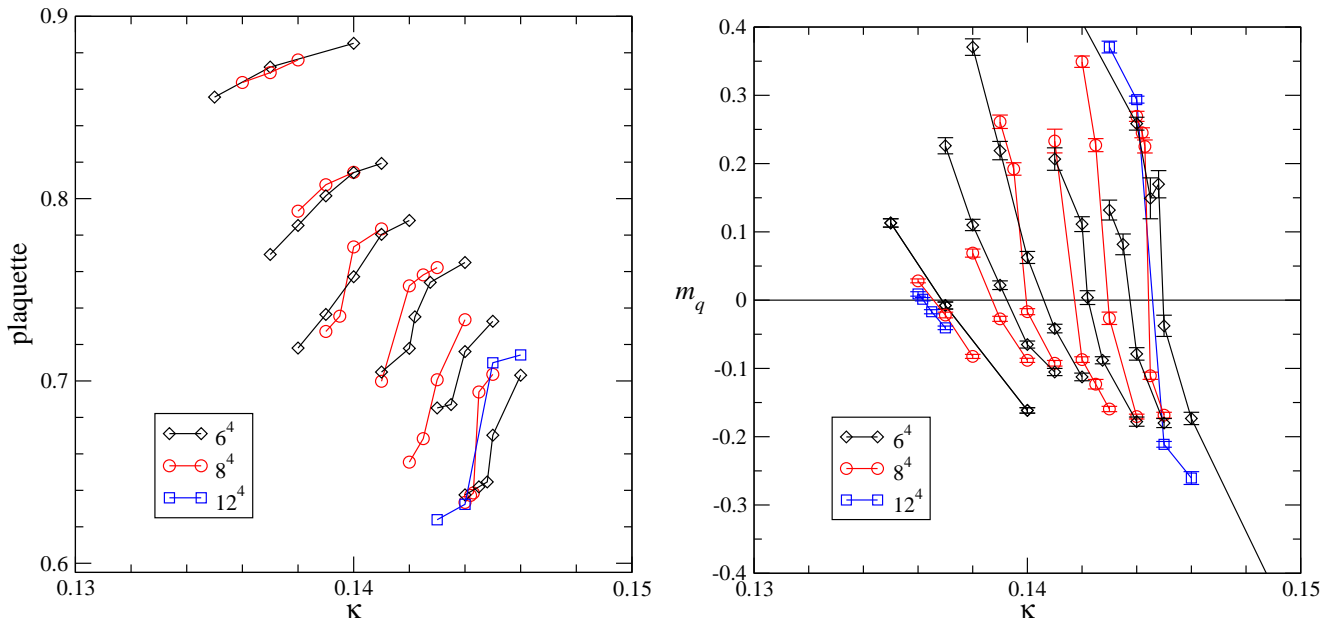

\begin{center}
\includegraphics[width=0.48\textwidth,clip]{plaq_scan_bqs.eps}
\hskip .02\columnwidth
\includegraphics[width=0.48\textwidth,clip]{mq0.eps}
\end{center}
\caption{Fixed-$\beta$ scans of the average plaquette (left) and the AWI mass (right) on three different volumes.
From right to left, the successive groups are for $\beta=1.2$, 1.25,
1.3, 1.35, 1.4, and 1.5.
For $\beta\le1.35$ there is a transition that strengthens with increased
volume.
\label{fig:ps}}
\end{figure*}
%%%%%%%%%%%%%%%%%%%%%%%%%%%%%%%%%%%%%%%%%%%%%%%%%%%%%%%%%%%%%%%%%%%%%

In order to measure the SF coupling, we had to map out the $\kappa_c$ line.
As discussed above, we determine $\kappa_c$ by demanding $m_q=0$ at fixed $\beta$; this is possible only for sufficiently weak coupling (large $\beta$---see Fig.~\ref{fig:bk}).
The $\kappa_c$ line meets at $\beta=\beta^*$  a line of first order transition at which the AWI quark mass
jumps discontinuously from a positive to a negative value; this makes it impossible
to define $\kappa_c$ for $\beta<\beta^*$.
On the first-order line, the discontinuity in $m_q$, like that in the plaquette, varies with $\beta$ and appears to vanish at $\beta^*$ (see Fig.~\ref{fig:ps});
this makes the meeting of the two lines a critical point.
This is similar to what was reported in Refs.~\cite{Catterall:2008qk,Hietanen:2008mr}.

\begin{figure}
\begin{center}
\includegraphics[width=\columnwidth,clip]{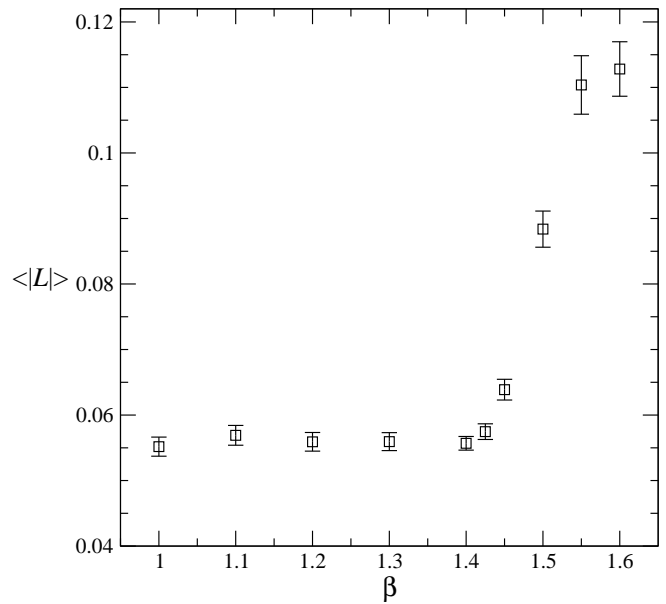}
\end{center}
\caption{Scan of the Polyakov loop average $\langle|L|\rangle$ at $\kappa=0.135$ on volume $6^4$.
\label{fig:kapsweep}}
\end{figure}

In finite volume, another line of transitions
separating the strong-coupling confining phase from a deconfined phase begins at or near the meeting point
and runs out towards $\kappa=0$.
If the spatial volume were to be made large, this would be the finite-temperature confinement transition.
The adjoint fermions leave the global Z(2) center symmetry unbroken, so the finite-temperature transition can be an Ising-like second order transition like that of the pure gauge theory.
We have investigated this transition at one value of $\kappa$ below the critical point and found it to be a continuous transition for volume $6^4$ (Fig.~\ref{fig:kapsweep}), quite different from the jumps seen in the fixed-$\beta$ scans of Fig.~\ref{fig:ps}. 
The simplest scenario is to suppose a second-order phase boundary that stretches from $\kappa=0$ to the critical point at $(\beta^*,\kappa^*)$.
%%%%%%%%%%%%%%%%%%%%%%%%%%%%%%%%%%%%%%%%%%%%%%%%%%%%%%%%%%%%%%%%%%%%%
%%%%%%%%%%%%%%%%%%%%%%%%%%%%%%%%%%%%%%%%%%%%%%%%%%%%%%%%%%%%%%%%%%%%%

%%%%%%%%%%%%%%%%%%%%%%%%%%%%%%%%%%%%%%%%%%%%%%%%%%%%%%%%%%%%%%%%%%%%%
\section{Running gauge coupling \label{sec:SF}}
%%%%%%%%%%%%%%%%%%%%%%%%%%%%%%%%%%%%%%%%%%%%%%%%%%%%%%%%%%%%%%%%%%%%%

\begin{figure}
\begin{center}
\includegraphics[width=\columnwidth,clip]{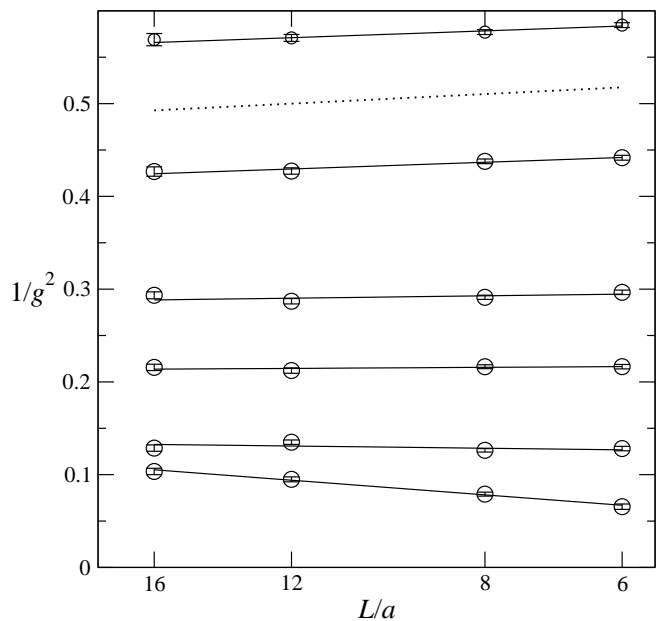}
\end{center}
\caption{SF coupling $1/g^2$ \textit{vs.}~$L/a$ (plotted on a logarithmic scale)
for the four volumes studied, at (from the top) $\beta=3.0$, 2.5, 2.0, 1.75,  1.5, and~1.4.
The lines through the data points are fits to the
data at each $\beta$ of the form $1/g^2(L)=a+b\log (L/a)$.
The dotted line has the slope $2b_1/(16\pi^2)$ as given by the lowest-order beta function,
Eq.~(\protect{\ref{eq:2loopbeta2}}).
\label{fig:1g2}}
\end{figure}

Our SF calculations were performed along the $\kappa_c$ line.
A summary of the data collected is  shown in Table~\ref{tab:runs}.
The measured SF couplings are tabulated in Table~\ref{tab:kg2} and plotted
(for some values of $\beta$) in Fig.~\ref{fig:1g2}. The  logarithmic
variation of $1/g^2$ with $L$ is characteristic of a slowly-running coupling.
The transition from positive to negative slope as the bare coupling
$\beta$ decreases is our first piece of evidence for the existence of an IRFP.

%%%%%%%%%%%%%%%%%%%%%%%%%%%%%%%%%%%%%%%%%%%%%%%%%%%%%%%%%%%%%%%%%%%%%
\begin{table*}
\caption{Schr\"odinger functional couplings $1/g^2$ from this study.}
\begin{center}
\begin{ruledtabular}
\begin{tabular}{ddrrrr}
\beta & \kappa_c &\multicolumn{4}{c}{$1/g^2$}\\
\cline{3-6}
&&              $L=6a$    & $L=8a$    & $L=12a$   & $L=16a$  \\
\hline
3.0 & 0.12682 &  0.5846(27) & 0.5771(26) & 0.5708(37)  & 0.5690(66) \\
2.5 & 0.1276 &  0.4417(27) & 0.4378(25) & 0.4273(34) & 0.4268(50)  \\
2.453 & 0.12766 & 0.4284(24)  &  \hfil-- &  0.4178(50) &   \hfil--  \\
2.445 & 0.12769 & 0.4305(27) & 0.4170(37) & \hfil-- &   \hfil--  \\
2.0 & 0.12967 & 0.2966(24) & 0.2912(23) & 0.2870(29) & 0.2934(37)  \\
1.985 & 0.12279 & 0.2912(24)  &  \hfil-- & 0.2801(48)  &   \hfil--  \\
1.97 & 0.12991 & 0.2852(24) & 0.2833(32) & \hfil-- &   \hfil--  \\
1.75 & 0.13216 & 0.2165(24) & 0.2164(21) & 0.2122(29) & 0.2157(34)       \\
1.5 & 0.13617 & 0.1281(24) & 0.1263(19) & 0.1350(24) & 0.1287(34) \\
1.4 & 0.13824 & 0.0655(29) & 0.0790(21) & 0.0950(27) & 0.1035(34) \\
\end{tabular}
\end{ruledtabular}
\end{center}
\label{tab:kg2}
\end{table*}
%%%%%%%%%%%%%%%%%%%%%%%%%%%%%%%%%%%%%%%%%%%%%%%%%%%%%%%%%%%%%%%%%%%%%

%%%%%%%%%%%%%%%%%%%%%%%%%%%%%%%%%%%%%%%%%%%%%%%%%%%%%%%%%%%%%%%%%%%%%
%%%%%%%%%%%%%%%%%%%%%%%%%%%%%%%%%%%%%%%%%%%%%%%%%%%%%%%%%%%%%%%%%%%%%

%%%%%%%%%%%%%%%%%%%%%%%%%%%%%%%%%%%%%%%%%%%%%%%%%%%%%%%%%%%%%%%%%%%%%
\subsection{The DBF for $s=2$}
%%%%%%%%%%%%%%%%%%%%%%%%%%%%%%%%%%%%%%%%%%%%%%%%%%%%%%%%%%%%%%%%%%%%%

\begin{figure}
\begin{center}
\includegraphics[width=\columnwidth,clip]{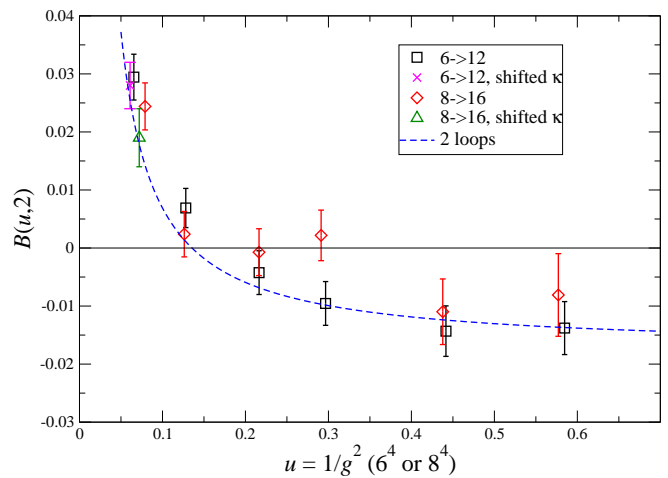}
\end{center}
\caption{Discrete beta function, \Eq{DBF}, for scale factor $s=2$ as a function of $1/g^2$ measured on the
smaller lattice.
Two values of $L$ are shown for the smaller lattice, $L=6$ and 8.
The pairs of points correspond to calculations carried out at (right to left)
$\beta=3.0$, 2.5, 2.0, 1.75,  1.5, and~1.4, as shown in Fig.~\ref{fig:1g2}, plus points calculated at a shifted $\kappa$ value at $\beta=1.4$.
The dashed line is the two-loop result.
The shifted $\kappa$ value is discussed in the Appendix.
\label{fig:SU2DBF2}}
\end{figure}

The data can be combined in various ways.
The most direct is to plot the $s=2$ DBF
for each of two values of $L/a$, 6 and 8.
This is shown in Fig.~\ref{fig:SU2DBF2}; it is essentially a comparison of the
DBF between two lattice spacings.
The data for $L/a= 6$ and 8 give a picture of the DBF with only weak dependence
on the lattice spacing.
Remarkably, the numerical result tracks the two-loop DBF rather closely.
We will estimate $g_*$, the location of the zero, in our more extensive
analysis below.

At $\beta=1.4$ we found that $\kappa_c$ was strongly dependent on volume.
We therefore calculated the SF coupling at the value of $\kappa_c$ appropriate
to $L=12a$ and then at a shifted value of $\kappa$.
We present a comparison of the two cases in the Appendix.
The shifted $\kappa$ value gives a separate determination of the DBF at
$\beta=1.4$, which is included in Fig.~\ref{fig:SU2DBF2}.
One sees that the shift in $\kappa$ leads to a change in the DBF that is
less than the statistical error bar.

%%%%%%%%%%%%%%%%%%%%%%%%%%%%%%%%%%%%%%%%%%%%%%%%%%%%%%%%%%%%%%%%%%%%%
%%%%%%%%%%%%%%%%%%%%%%%%%%%%%%%%%%%%%%%%%%%%%%%%%%%%%%%%%%%%%%%%%%%%%

%%%%%%%%%%%%%%%%%%%%%%%%%%%%%%%%%%%%%%%%%%%%%%%%%%%%%%%%%%%%%%%%%%%%%
\subsection{Finding $g_*$ and estimating systematic error \label{2fit}}
%%%%%%%%%%%%%%%%%%%%%%%%%%%%%%%%%%%%%%%%%%%%%%%%%%%%%%%%%%%%%%%%%%%%%

As can be seen from Fig.~\ref{fig:1g2}, the spread of values of $1/g^2$
for different volumes at fixed bare coupling $\beta$ is typically much smaller
than the  difference between values obtained at different $\beta$'s.
This motivates us to analyze the data in two stages.
In the first stage, the data set at each $\beta$ is treated
as an independent fitting problem.  The outcome is the value of the DBF
at some reference value of $1/g^2$ reachable at that $\beta$.
In the second stage, the DBF's from all $\beta$'s are combined
to obtain an estimate of $g_*$, the location of the IRFP.
The variety of fits studied gives us a handle on the systematic error
in $g_*$.

The data plotted in Fig.~\ref{fig:1g2} are
evidently linear in $\log(L/a)$ for fixed $\beta$.
To study deviations from linearity
we try fitting four different functions to $u\equiv1/g^2(L)$:
\begin{subequations}\label{grp:fits}
\begin{align}
u&= a+b\log x, \label{fit:a}\\
u&=a+b\log x+c(\log x)^2,\label{fit:b}\\
u&= a+b\log x+c/x,\label{fit:c}\\
u&=a+b\log x+c/x^2.\label{fit:d}
\end{align}
\end{subequations}
In these formulas, $x=L/8a$.  The simplest fit, \Eq{fit:a}, assumes linearity
in $\log(L/a)$ and no discretization errors.  The results of this fit
at each bare coupling were plotted in Fig.~\ref{fig:1g2}.
In \Eq{fit:b} we have added
the next term from \Eq{eq:fit1}, reflecting subleading continuum running.
In the last two fit functions we include instead a term that accounts for
discretization errors.  In \Eq{fit:c} we assume that the leading lattice
artifacts are linear in $a/L$,
whereas in \Eq{fit:b} they are assumed to be quadratic.

In the continuum limit, $a/L\to 0$, fits~(\ref{fit:c}) and~(\ref{fit:d})
both reduce to the simplest fit, \Eq{fit:a}.  The fit parameter $a$
is thus interpreted as $1/g^2(L=8a)$, while $b$ is the estimate $R(g^2)$
for the beta function $\tbeta(1/g^2)$ at the same $L$.
The parameters $a$ and $b$ have a similar interpretation
for fit~(\ref{fit:b}) as well.
In particular, since $\tbeta$ for $1/g^2(L=8a)$
is $\partial u/\partial \log s$ evaluated at $s=1$, this is again $b$.

%%%%%%%%%%%%%%%%%%%%%%%%%%%%%%%%%%%%%%%%%%%%%%%%%%%%%%%%%%%%%%%%%%%%%
\begin{figure*}
\begin{center}

\includegraphics*[width=.48\textwidth]{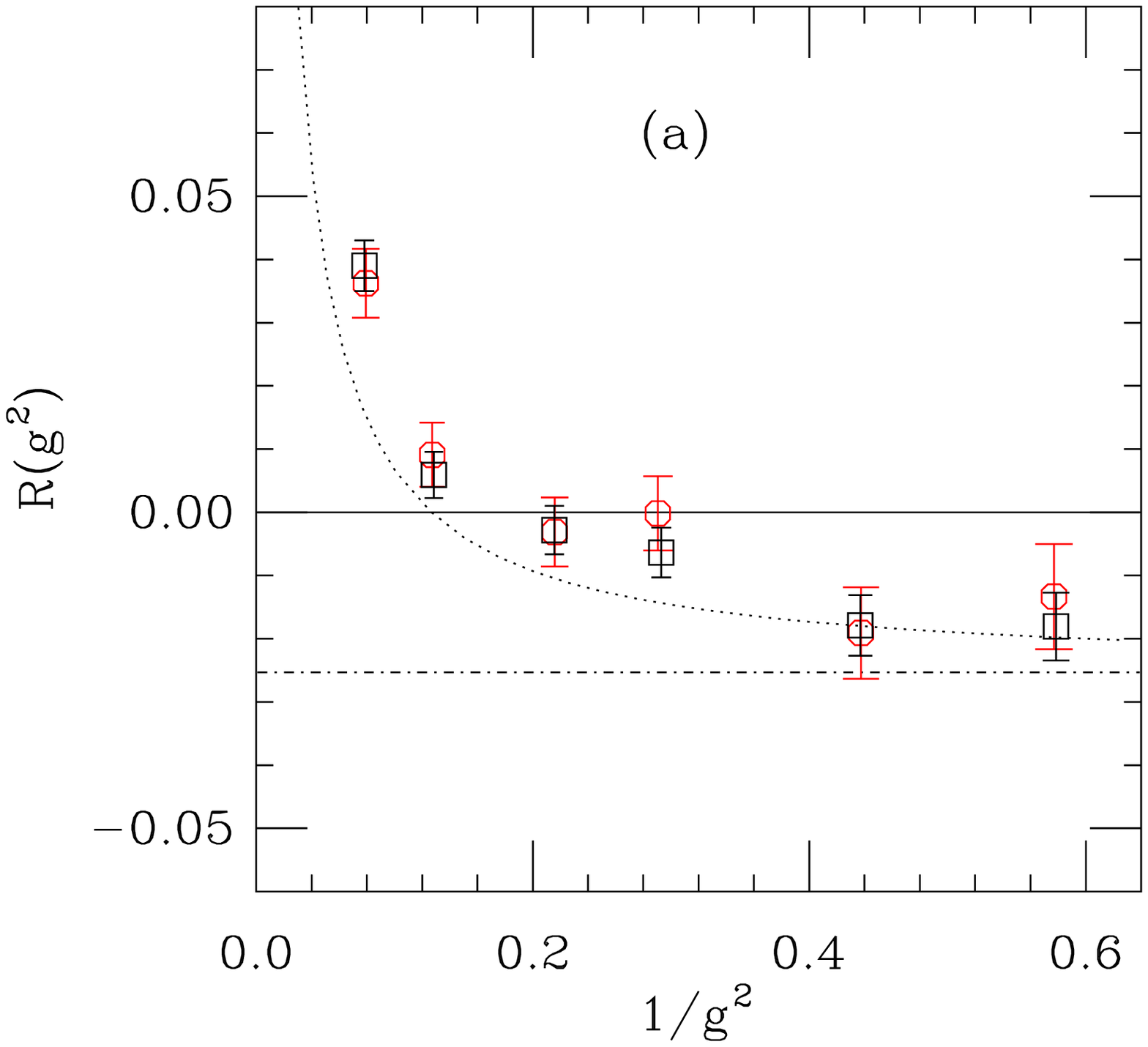}
\hspace{.02\textwidth}
\includegraphics*[width=.48\textwidth]{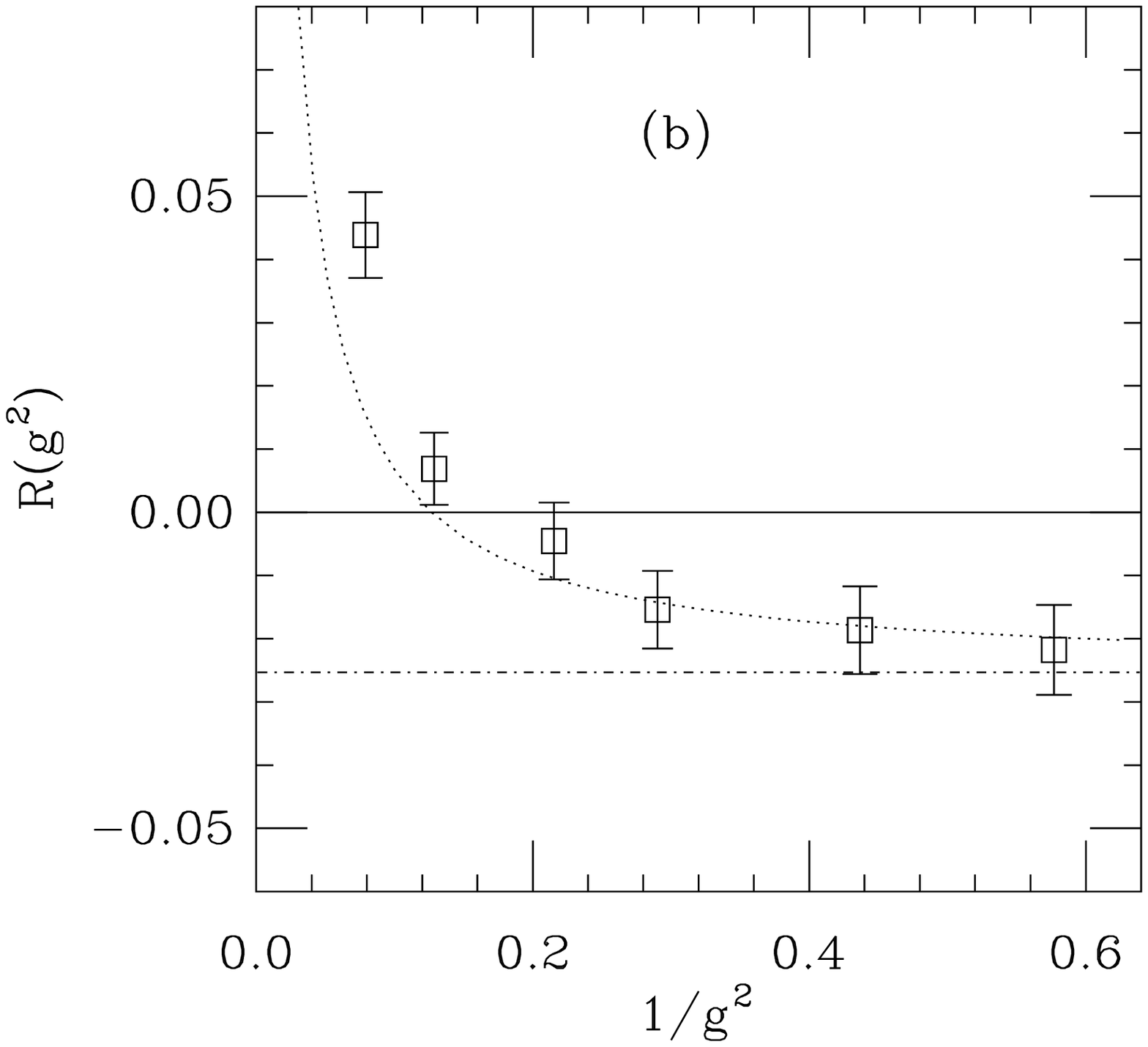}

\vspace*{5ex}

\includegraphics*[width=.48\textwidth]{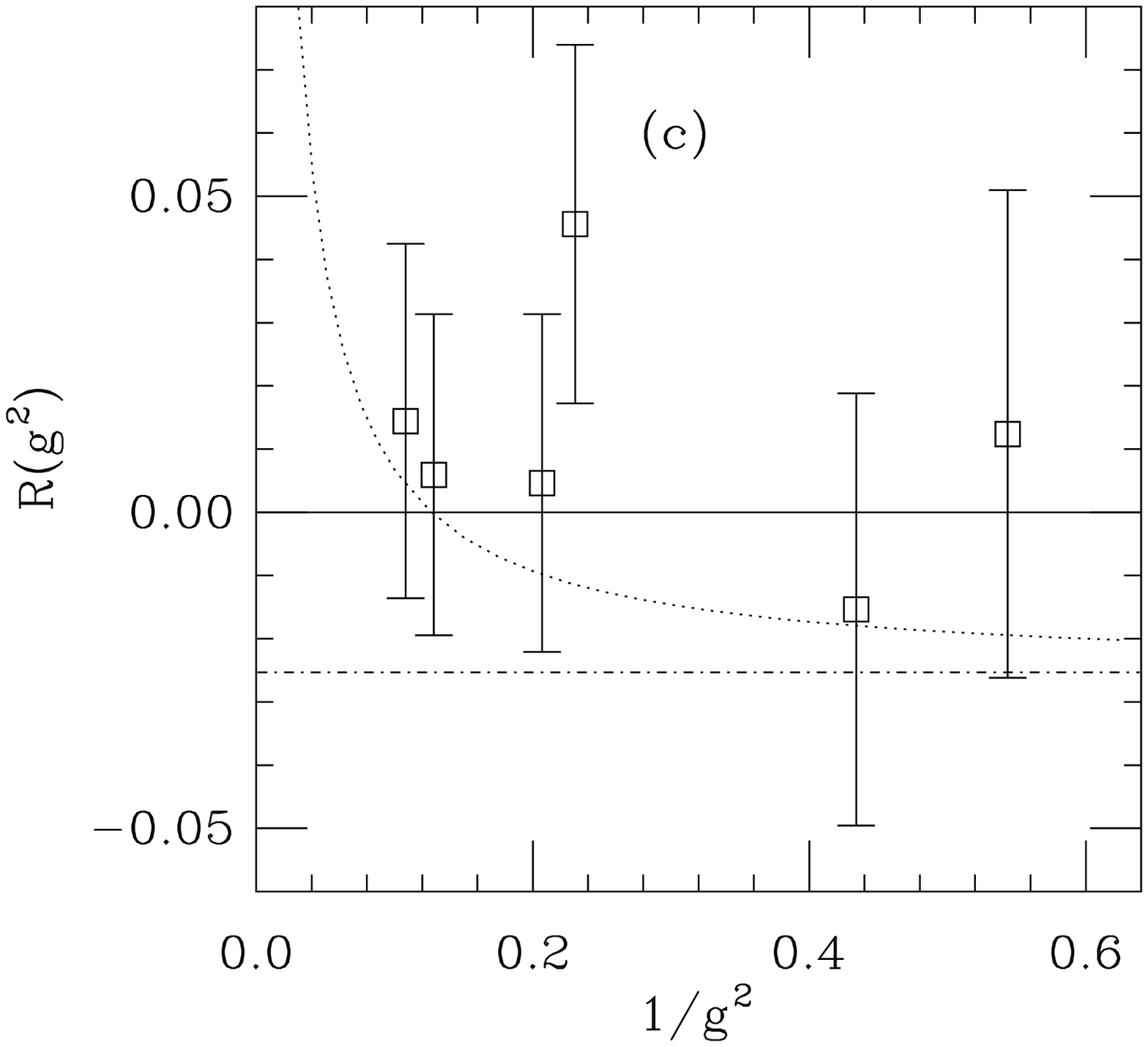}
\hspace{.02\textwidth}
\includegraphics*[width=.48\textwidth]{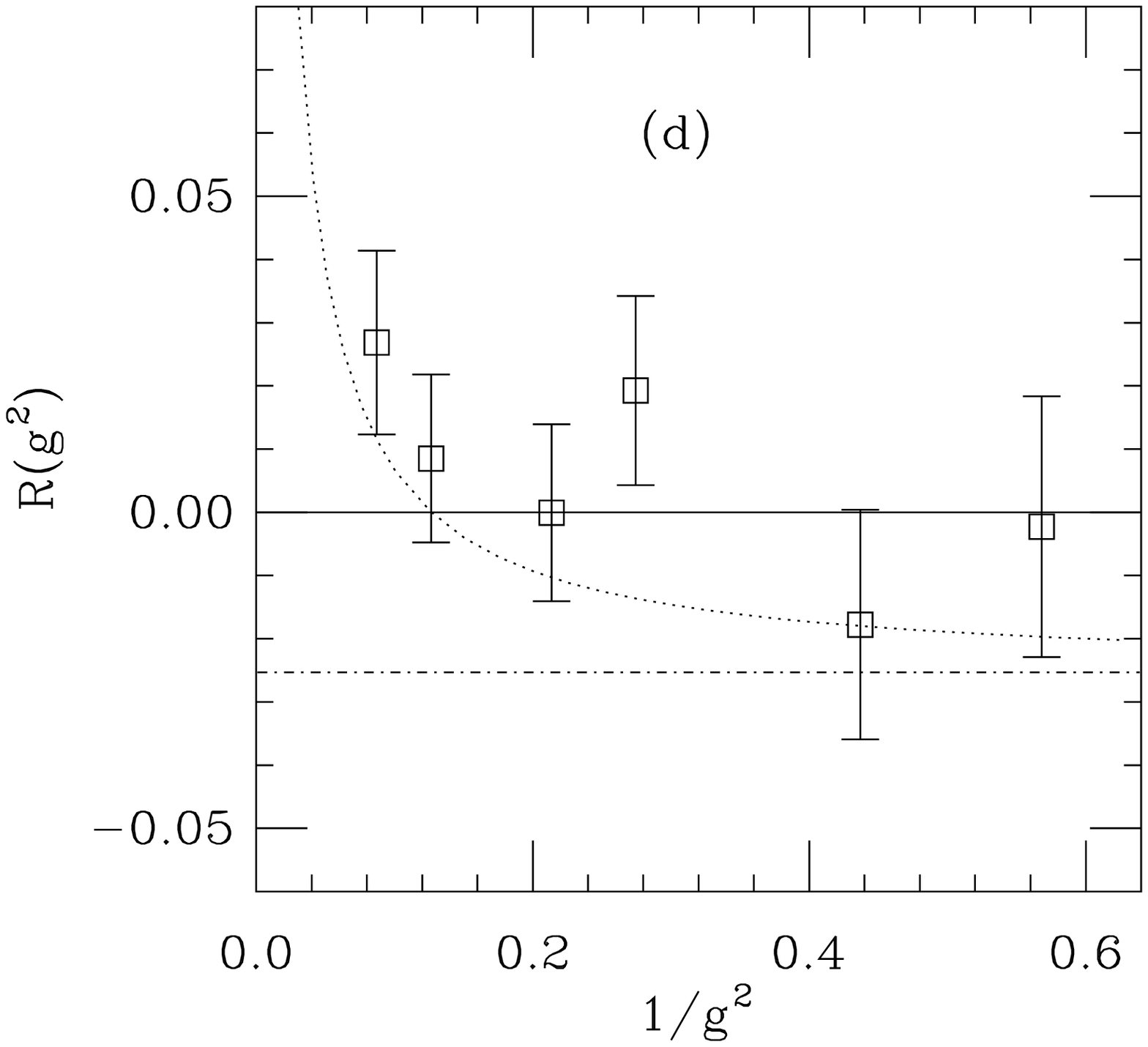}
\vspace*{0ex}
\end{center}
\caption{Estimate $R(g^2)$ for the beta function $\tbeta(1/g^2)$ as extracted from the fits listed in \Eq{grp:fits}.
Each panel shows results of a different fit,
as described more fully in the text:
(a) Squares show fits to $a+b\log x$ using all four volumes;
circles, using $L/a=8$, 12, 16 only.
(b) Fits to $a+b\log x+c(\log x)^2$.
(c) Fits to $a+b\log x+c/x$.
(d) Fits to $a+b\log x+c/x^2$.
Plotted curves are the one-loop (dashed line)
and two-loop (dotted line) beta functions.
\label{fig:bfndatam}}
\end{figure*}
%%%%%%%%%%%%%%%%%%%%%%%%%%%%%%%%%%%%%%%%%%%%%%%%%%%%%%%%%%%%%%%%%%%%%

We plot the fit parameter $b$, which gives the beta function, versus $a$,
the inverse coupling, for each fit type in Fig.~\ref{fig:bfndatam}.
The squares show results of fits using all four volumes,
$L/a=6$, 8, 12, 16.
These fits have one degree of freedom, except for the simplest fit~(\ref{fit:a})
which has two degrees of freedom.  We have included in panel~(a) another fit
to \Eq{fit:a} where only the three largest volumes ($L/a=8$, 12, 16) are kept.
This fit has one degree of freedom.

For almost all beta values, all of these fits produce $\chi^2/$dof near unity;
the exception is $\beta=1.5$, where all fits give $\chi^2\simeq 6$.
The five fit types give
results that are consistent with each other as well as with the two-loop
beta function.  We remark that all fits have good $\chi^2/$dof at
the strongest bare coupling, $\beta=1.4$,
leaving no doubt that the beta function has crossed zero.

%%%%%%%%%%%%%%%%%%%%%%%%%%%%%%%%%%%%%%%%%%%%%%%%%%%%%%%%%%%%%%%%%%%%%
\begin{table*}
\caption{Linear fits to the beta functions plotted in Fig.~\ref{fig:bfndatam}.
Each beta function (resulting from the fitting procedure listed in the table)
is a fit to $R(g^2)=A+B/g^2$, and the fit gives an estimate of the zero of $R$
at $g_*$. }
\begin{center}
\begin{ruledtabular}
\begin{tabular}{lddd}
fit type giving beta function & A & B &   1/g^2_*  \\
\hline
(a) $a+b\log x$, all volumes     & 0.015(5) & -0.08(2)      &  0.20(3)  \\
(a) $a+b\log x$, $L/a=8$, 12, 16 & 0.019(7) & -0.08(3)      &  0.23(4)  \\
(b) $a+b\log x+c(\log x)^2$      &  0.015(8) & -0.09(3)     &  0.17(4)  \\
(c) $a+b\log x+c/x$              &   0.027(34) & -0.07(14)  &  0.40(41) \\
(d) $a+b\log x +c/x^2$           & 0.020(18) & -0.07(7)     &  0.31(13) \\
\end{tabular}
\end{ruledtabular}
\end{center}
\label{tab:gstarfits}
\end{table*}
%%%%%%%%%%%%%%%%%%%%%%%%%%%%%%%%%%%%%%%%%%%%%%%%%%%%%%%%%%%%%%%%%%%%%

Our next task is the determination of $g_*$,
the value of the running coupling where the DBF vanishes.
For each fit type, we can locate the zero by
a linear fit to the points in its figure.
In all cases, we get good $\chi^2$  after dropping the points at the smallest
($\beta=1.4$) and largest ($\beta=3.0$) couplings.
The final results are shown in Table~\ref{tab:gstarfits}.
One observes that all the results are mutually consistent.

The differences among the estimates quoted
in Table~\ref{tab:gstarfits} for $1/g_*^2$ reflect our systematic uncertainties,
to which we now turn.
Unlike in QCD, where all sources of systematic error
are well under control, in the case of a (nearly) conformal theory
the systematic error is poorly understood.  Indeed, an important conclusion
from Fig.~\ref{fig:bfndatam} is that our data do not allow us to sort out
discretization errors from subleading continuum corrections.  With this in mind,
we estimate the systematic error by keeping a subset of the five fit
types that represents both options.

First, we obviously keep the simplest,
linear fit~(\ref{fit:a}) on all four volumes.  To account for the possibility
of continuum corrections, we include fit~(\ref{fit:b}).

It remains to select a fit that represents the possible
discretization errors.  Here we face a difficulty.
As can be seen in Fig.~\ref{fig:bfndatam} and in Table~\ref{tab:gstarfits},
fits~(\ref{fit:c}) and~(\ref{fit:d}), which both include a term for lattice artifacts,
give significantly larger error bars than the other fits.
This shows that our data do not resolve 
$\log x$ from $1/x^2$ (and, even more so, from $1/x$).
We stress that our clover action with the nHYP links generally shows much
smaller discretization errors than the simple Wilson action.
Nonetheless, since we keep the clover coefficient at its tree-level value
of 1, some residual linear dependence on $a/L$ could survive.

Luckily, we have yet another fit type that is sensitive
to the discretization errors and, at the same time,
produces much tighter uncertainties in $g_*$.
This is the simple fit~(\ref{fit:a})
in which we drop the smallest volume, $L=6a$.  Since the smallest volume
necessarily contains the largest discretization errors, dropping it must
give us a result that is closer to the continuum limit.
The advantage of dropping the smallest volume in the linear fit
over fits~(\ref{fit:c})
and~(\ref{fit:d}) is that there is no need to postulate anything
about the concrete form of the discretization errors; in particular,
we do not have to assume anything about the relative size of
$a/L$ and $(a/L)^2$ errors.  

Disregarding fits~(\ref{fit:c}) and~(\ref{fit:d}) while keeping
the other three, we finally conclude that
\bee
\frac{1}{g_*^2} =0.20(4)(3) ,
\label{invgstar}
\ee
where the first error is statistical and the second is systematic,
representing the spread of the mean values
of the three selected fit types.  It follows that
\bee
g_*^2 = 5.0^{+2.7}_{-1.3}\ ,
\label{gstar}
\ee
where we have combined the systematic and statistical errors linearly.%
\footnote{Examination of our graphs shows that most of the systematic error is due to
one data point for $1/g^2$, namely that at $\beta=2.0$ for $L=16a$.
This is responsible for the high point in the $8\to16$ DBF plotted in
Fig.~\ref{fig:SU2DBF2} at $u\simeq0.3$.}

The derivative of the beta function at the fixed point is a universal quantity.
In our linear fits, reported in Table~\ref{tab:gstarfits},
this is just the fit parameter $B$ .  We conclude
\bee
\left.\frac{d\tbeta}{du} \right|_{u=1/g_*^2} = -0.08(3).
\ee
This time the error is entirely statistical.  The three fit types
we have kept produce essentially the same result, so the systematic
error is negligible.  This translates into an exponent
\bee
y_g = \left.\frac{d\beta(g^2)}{dg^2} \right|_{g_*^2}
= -\frac{1}{2}B = 0.040(15).
\ee
The positive sign indicates infrared irrelevancy of the gauge coupling.

%%%%%%%%%%%%%%%%%%%%%%%%%%%%%%%%%%%%%%%%%%%%%%%%%%%%%%%%%%%%%%%%%%%%%
\begin{figure*}
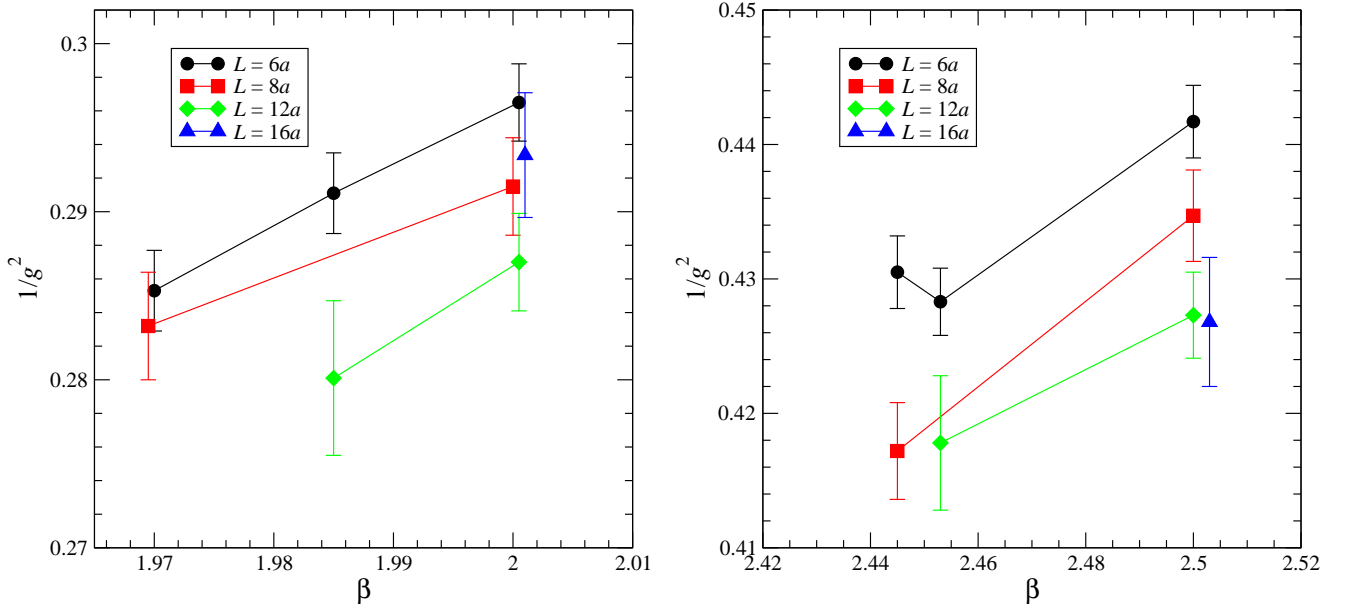

\begin{center}
\includegraphics[width=0.48\textwidth,clip]{1g2_vs_b_2.0_bqs.eps}
\hskip .02\columnwidth
\includegraphics[width=0.48\textwidth,clip]{1g2_vs_b_2.5_bqs.eps}
\end{center}
\caption{$1/g^2$ near  $\beta=2.0$ (left) and $\beta=2.5$ (right).
Points at $\beta=2.0$ and $\beta=2.5$ are slightly displaced for clarity.
\label{fig:1g22.02.5}}
\end{figure*}
%%%%%%%%%%%%%%%%%%%%%%%%%%%%%%%%%%%%%%%%%%%%%%%%%%%%%%%%%%%%%%%%%%%%%

%%%%%%%%%%%%%%%%%%%%%%%%%%%%%%%%%%%%%%%%%%%%%%%%%%%%%%%%%%%%%%%%%%%%%%%%%%%%%%
\subsection{Attempt to disentangle continuum running from lattice artifacts}
%%%%%%%%%%%%%%%%%%%%%%%%%%%%%%%%%%%%%%%%%%%%%%%%%%%%%%%%%%%%%%%%%%%%%

Here we describe an attempt at a more traditional SF analysis,
as described in Sec.~\ref{sec:method}. As we have seen above,
data taken at the same value of bare parameters, but at several values of $L$,
will show a combination of true running and lattice artifacts.
In principle, these effects can be separated: One
adjusts the bare parameters in the simulation to match the
SF coupling on two different-sized small lattices, and
then increases the lattice sizes by the same scale factor $s$.
A comparison of the DBF's obtained will show only lattice artifacts.
This differs from our plot of the DBF for $s=2$, Fig.~\ref{fig:SU2DBF2},
in that here we calculate $B(u,s)$ for fixed $u=1/g^2$
with two different lattice spacings;
in the plot we kept $(\beta,\kappa)$ fixed,
and hence the lattice spacing, between lattices with $L=6a$ and $L=8a$.
It basically amounts to using lattice data at several couplings as a substitute for our
fitting functions Eq.~\ref{fit:c} or \ref{fit:d}.

\begin{figure*}
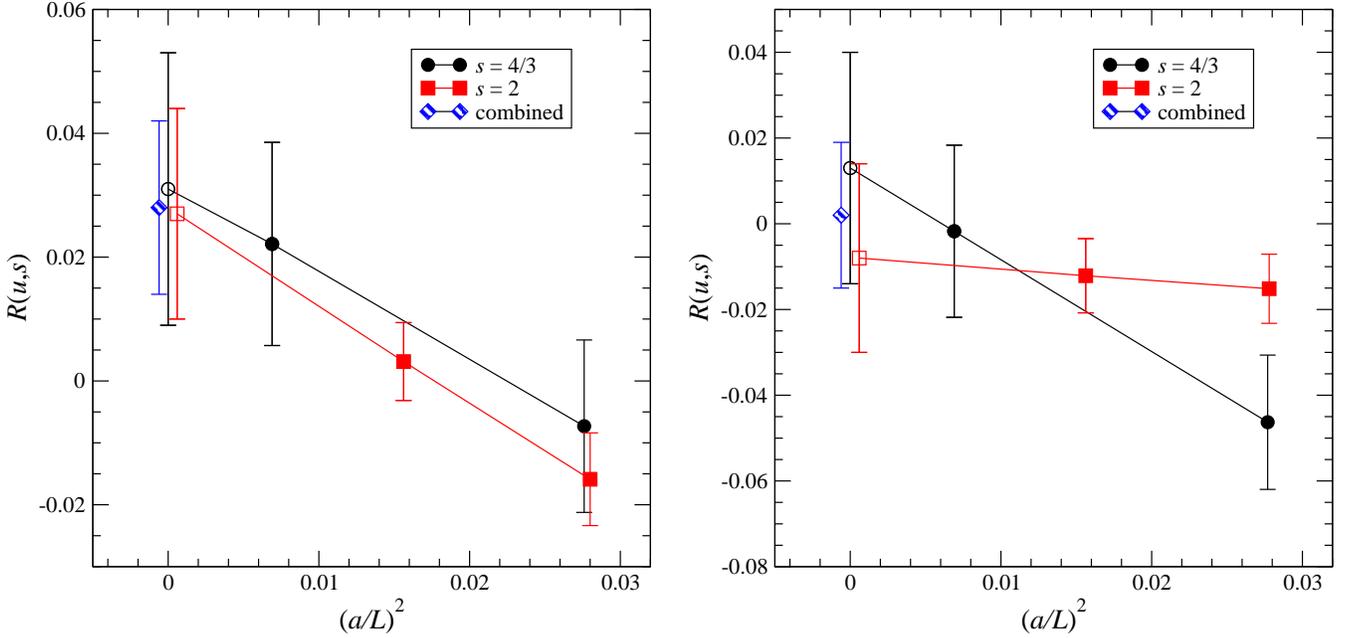

\begin{center}
\includegraphics[width=0.48\textwidth,clip]{dbf_l_2.0f_bqs.eps}
\hskip .02\textwidth
\includegraphics[width=0.48\textwidth,clip]{dbf_l_2.5f_bqs.eps}
\end{center}
\caption{Discrete beta functions near $\beta=2.0$ (left)
  and $\beta=2.5$ (right), where the smaller lattices'
$g^2$ values are matched by slightly varying the bare parameters.
The points near $(a/L)^2=0$ are the results of fits, described in the text:
the circle is the extrapolation of $R(u,4/3)$, the square is the extrapolation
of $R(u,2)$, and the diamond is a fit to all four points.
Some points have been displaced slightly for clarity.
\label{fig:dbf2.02.5}}
\end{figure*}

We attempted such a match near two values of the bare coupling,
$\beta=2.0$ and 2.5.  In Figs.~\ref{fig:1g22.02.5} and~\ref{fig:dbf2.02.5},
we calculate $B(u,2)$ using $L=8a$ and~$16a$
at one bare parameter value $\beta$, and then use $L'=6a'$ and~$12a'$ at
$\beta'$; the relation between $\beta$ and $\beta'$ is
$u(L'=6a',\beta')=u(L=8a,\beta)$.
We may then declare that $L'=L$ in physical units.
We can also do the same exercise for $s=4/3$ by starting with $L=12a$
and~$16a$ at $\beta$, matching to $L'=6a'$ and~$8a'$ at $\beta'$.
The data for $1/g^2$ are shown in Fig.~\ref{fig:1g22.02.5}.
It is apparent that matching the couplings, for example
$u(L'=6a',\beta')$ and $u(L=8a,\beta)$, can only be accomplished within
large error bars.

Fig.~\ref{fig:dbf2.02.5} shows the rescaled DBF $R(u,s)$ for scale
factors $s=4/3$ and~2.  We plot $R$ against $(a/L)^2$,
and attempt to fit the data to a linear dependence in $(a/L)^2$.
We can do it separately for each value of $s$, or,
following the discussion in Sec.~\ref{sec:method},
we can fit the rescaled DBF to a common line.
The results of those fits are shown in the two figures.
The fits of all the data give $\chi^2=0.4$ and~3.2
for two degrees of freedom at $\beta=2.0$ and 2.5, respectively.

At $\beta=2.5$, or $1/g^2 \simeq 0.42$, the continuum-extrapolated rescaled DBF
$R(u,s)$ is
0.013(27) for $s=4/3$, $-0.008(22)$ for $s=2$, and 0.002(17)
for the combined fit.  The two loop result is $-0.018$,
and our numerical result is consistent with it.

At $\beta=2.0$, or $1/g^2 \simeq 0.27$, the continuum-extrapolated
$R(u,s)$ is 0.031(22) for $s=4/3$, 0.027(17) for $s=2$,
and 0.024(14) for the combined fit. This is  $1.7\sigma$ away from zero.
The two-loop result is $-0.014$, about $2\sigma$ away.
The fits with many $L$'s using Eqs.~(\ref{fit:c})--(\ref{fit:d}),
which include {\em ans\"atze} for discretization errors,
also produced a positive DBF with a large uncertainty at $\beta=2.0$.
The SF coupling is  right on the edge of the value we quote in \Eq{invgstar} for $1/g_*^2$
from our analysis of many couplings.

Notice that the analysis in this subsection underestimates
the error in the DBF  since we have not included the uncertainty in $\beta'$
that arises from matching the SF couplings.
To make a definitive determination of $g_*^2$  using the extrapolation method  would require
repeating it at many couplings with significantly better statistics.
We conclude that this method is no better than fits to
Eqs.~(\ref{fit:c})--(\ref{fit:d}).
We have already argued that the other fits, Eqs.~(\ref{fit:a})--(\ref{fit:b}), 
give more reliable results.

%%%%%%%%%%%%%%%%%%%%%%%%%%%%%%%%%%%%%%%%%%%%%%%%%%%%%%%%%%%%%%%%%%%%%
%%%%%%%%%%%%%%%%%%%%%%%%%%%%%%%%%%%%%%%%%%%%%%%%%%%%%%%%%%%%%%%%%%%%%%%%%%%%%%

%%%%%%%%%%%%%%%%%%%%%%%%%%%%%%%%%%%%%%%%%%%%%%%%%%%%%%%%%%%%%%%%%%%%%
\section{Mass anomalous dimension \label{sec:zp}}
%%%%%%%%%%%%%%%%%%%%%%%%%%%%%%%%%%%%%%%%%%%%%%%%%%%%%%%%%%%%%%%%%%%%%

%%%%%%%%%%%%%%%%%%%%%%%%%%%%%%%%%%%%%%%%%%%%%%%%%%%%%%%%%%%%%%%%%%%%%
\begin{table}
\caption{Values of $Z_P$, the pseudoscalar renormalization constant}
\begin{center}
\begin{ruledtabular}
\begin{tabular}{ddrrrr}
\beta & \kappa_c &\multicolumn{4}{c}{$Z_P$}\\
\cline{3-6}
&&              $L=6a$    & $L=8a$    & $L=12a$   & $L=16a$  \\
\hline
3.0 & 0.12682 & 0.511(1) & 0.490(1) & 0.462(2) & 0.441(2)   \\
2.5 & 0.1276 & 0.484(1) & 0.456(1) & 0.423(1) & 0.402(1)  \\
2.0 & 0.12967 & 0.427(1) & 0.394(1) & 0.354(1) & 0.332(1) \\
1.75 & 0.13216 & 0.373(1) & 0.338(1) & 0.301(2) & 0.277(1) \\
1.5 & 0.13617 & 0.293(1) & 0.261(1) & 0.227(1) & 0.203(1)  \\
1.4 & 0.13824 & 0.250(3) & 0.219(2) & 0.188(1) & 0.173(2) \\
\end{tabular}
\end{ruledtabular}
\end{center}
\label{tab:zp}
\end{table}
%%%%%%%%%%%%%%%%%%%%%%%%%%%%%%%%%%%%%%%%%%%%%%%%%%%%%%%%%%%%%%%%%%%%%

Finally we turn to the mass anomalous dimension $\gamma_m$.
Our analysis will parallel that of Sec.~\ref{2fit} with the
difference that $\log Z_P(L)$ replaces $1/g^2(L)$.

\begin{figure}
\begin{center}
\includegraphics[width=\columnwidth,clip]{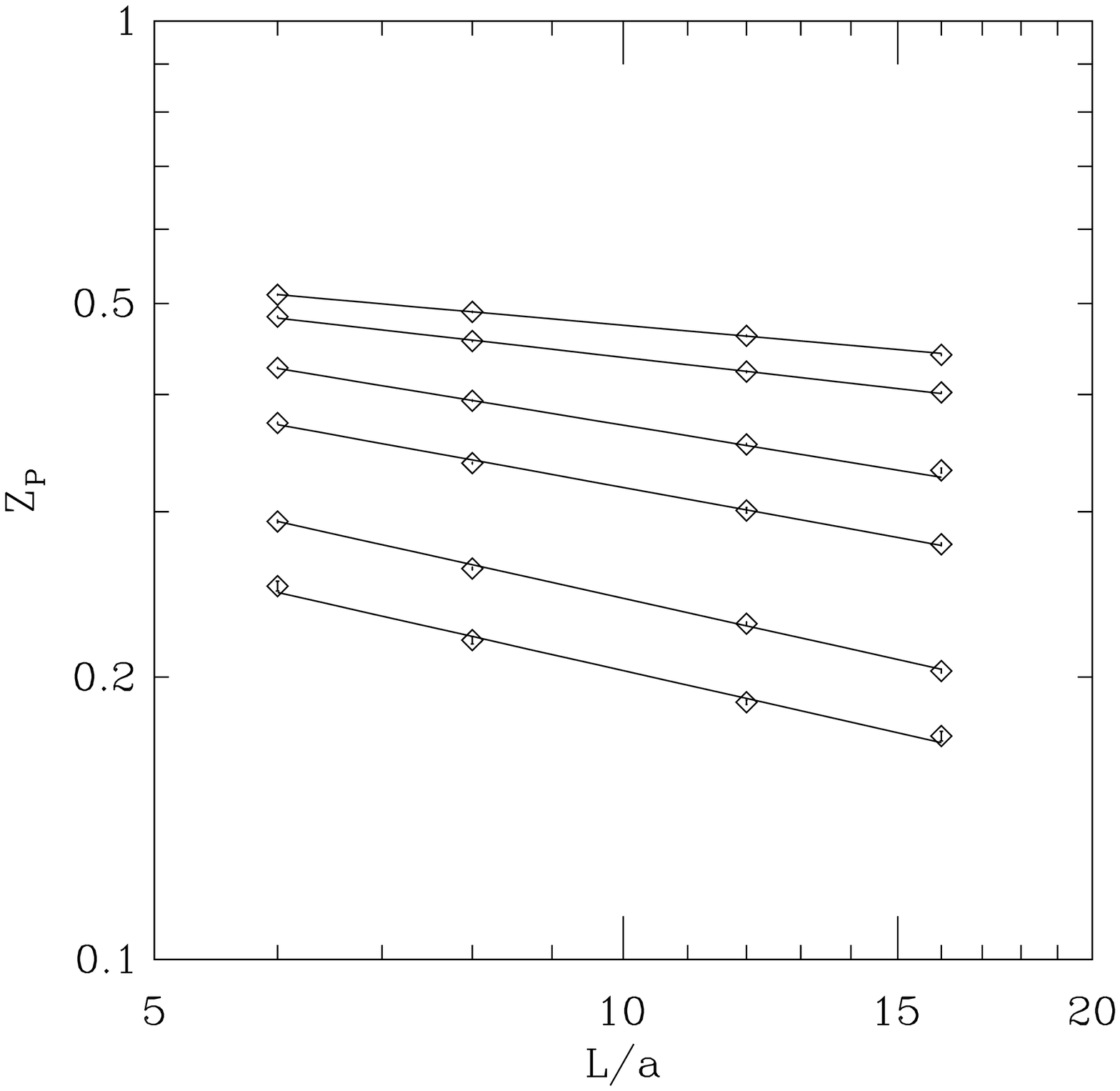}
\end{center}
\caption{Pseudoscalar renormalization constant $Z_P$.
From the top, data are from $\beta=3.0$, 2.5, 2.0, 1.75, 1.5 and~1.4.
Lines are fits to
$\log Z_P(L) = -\gamma_m\log (L/a) + \text{const}$ for each $\beta$.
\label{fig:zp}}
\end{figure}

Our data for $Z_P(L)$ are shown in Table~\ref{tab:zp} and plotted in
Fig.~\ref{fig:zp}.
The lines represent fits of the form of \Eq{eq:lgamma} to the data
at each bare parameter value.
The nearly straight-line behavior of the data
is indicative of a slowly running coupling.

%%%%%%%%%%%%%%%%%%%%%%%%%%%%%%%%%%%%%%%%%%%%%%%%%%%%%%%%%%%%%%%%%%%%%
%%%%%%%%%%%%%%%%%%%%%%%%%%%%%%%%%%%%%%%%%%%%%%%%%%%%%%%%%%%%%%%%%%%%%

\begin{figure}
\begin{center}
\includegraphics[width=\columnwidth,clip]{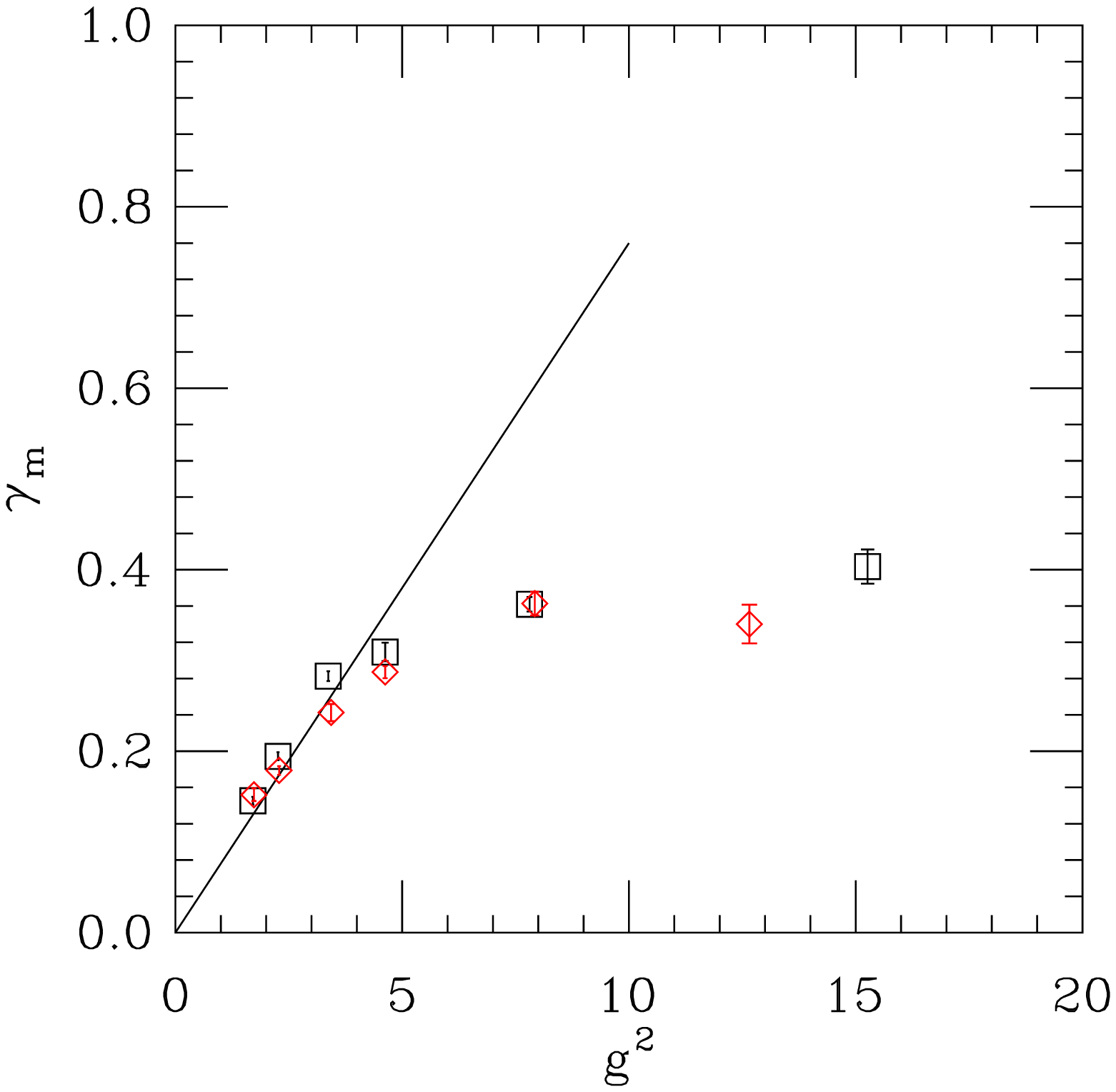}
\end{center}
\caption{Anomalous dimension $\gamma_m$  calculated from $\sigma_P(v,s=2)$ using  \Eq{eq:gamma}
from pairs of lattices: $L/a=6\to12$ (squares) and $L/a=8\to16$ (diamonds).
It is  plotted as a function of the SF coupling $g^2$ on the smaller lattice size.
Bare couplings range from $\beta=3$ on the left to 1.4 on the right, as in
Fig.~\ref{fig:zp}.
The line is the lowest-order perturbative result.
\label{fig:effgammasu2}}
\end{figure}

Fig.~\ref{fig:effgammasu2} shows the values of $\gamma_m$
extracted using Eqs.~(\ref{eq:sigma_p}) and~(\ref{eq:gamma})
from pairs of lattices with scale factor $s=2$.
We have plotted the data as a function of the SF coupling $g^2$ measured on the
smaller volume.  There does not seem to be a great deal of difference
between results from the two pairs, meaning that there is not much of a shift with lattice spacing.
The rightmost points are from the strongest coupling, $\beta=1.4$\@.
As is clear in our various determinations of the DBF, the SF coupling does run significantly at this value of $\beta$.
Note again that the $L=6a$ coupling is larger than the $L=8a$ coupling, indicative of the positive DBF.

The main feature of Fig.~\ref{fig:effgammasu2} is that the measured anomalous dimension $\gamma_m$ at
first follows closely the one-loop curve, but beyond $g^2\simeq 4$
it flattens out.

%%%%%%%%%%%%%%%%%%%%%%%%%%%%%%%%%%%%%%%%%%%%%%%%%%%%%%%%%%%%%%%%%%%%%
%%%%%%%%%%%%%%%%%%%%%%%%%%%%%%%%%%%%%%%%%%%%%%%%%%%%%%%%%%%%%%%%%%%%%

\begin{figure*}
\begin{center}

\includegraphics*[width=.48\textwidth]{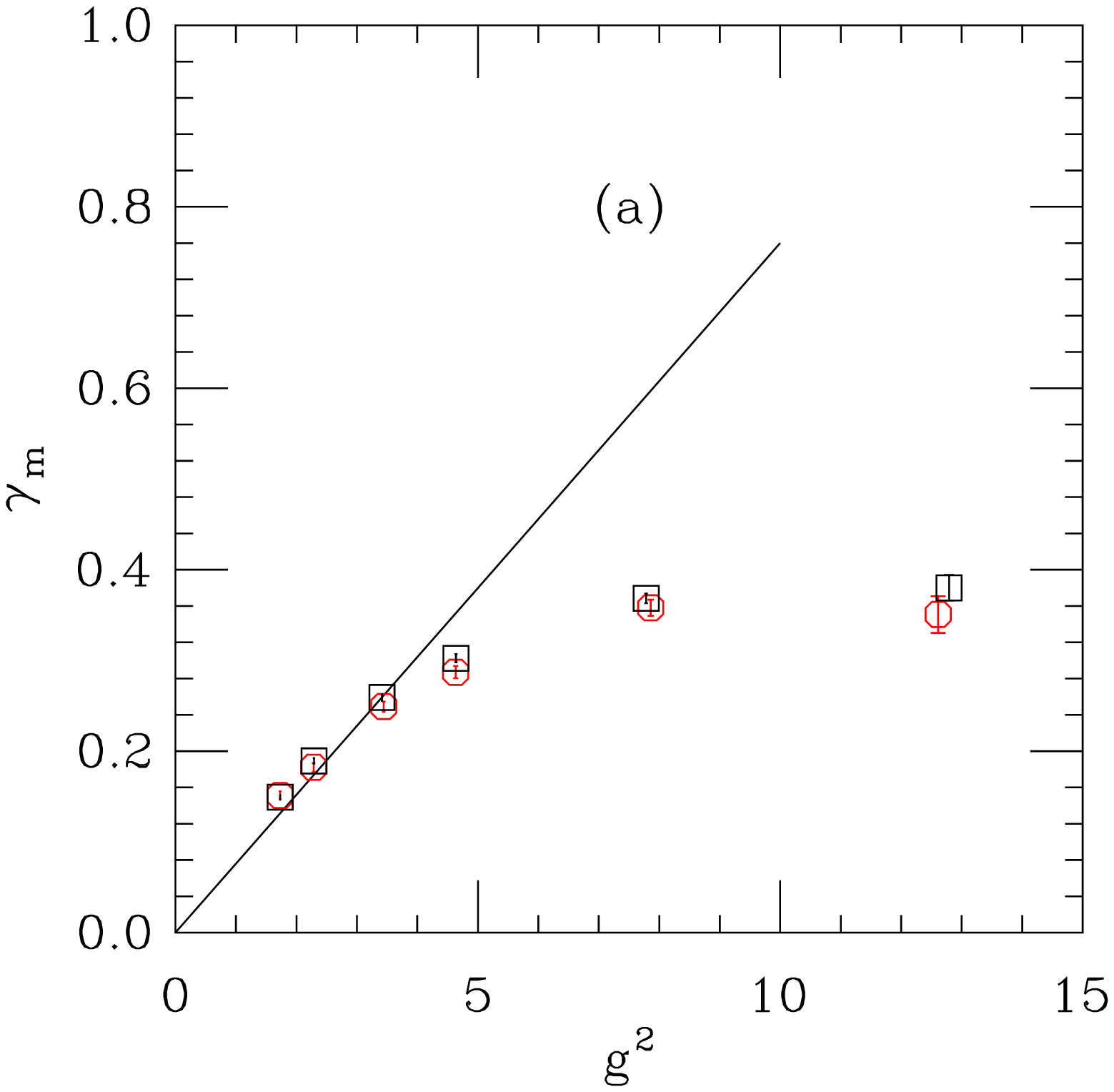}
\hspace{.02\textwidth}
\includegraphics*[width=.48\textwidth]{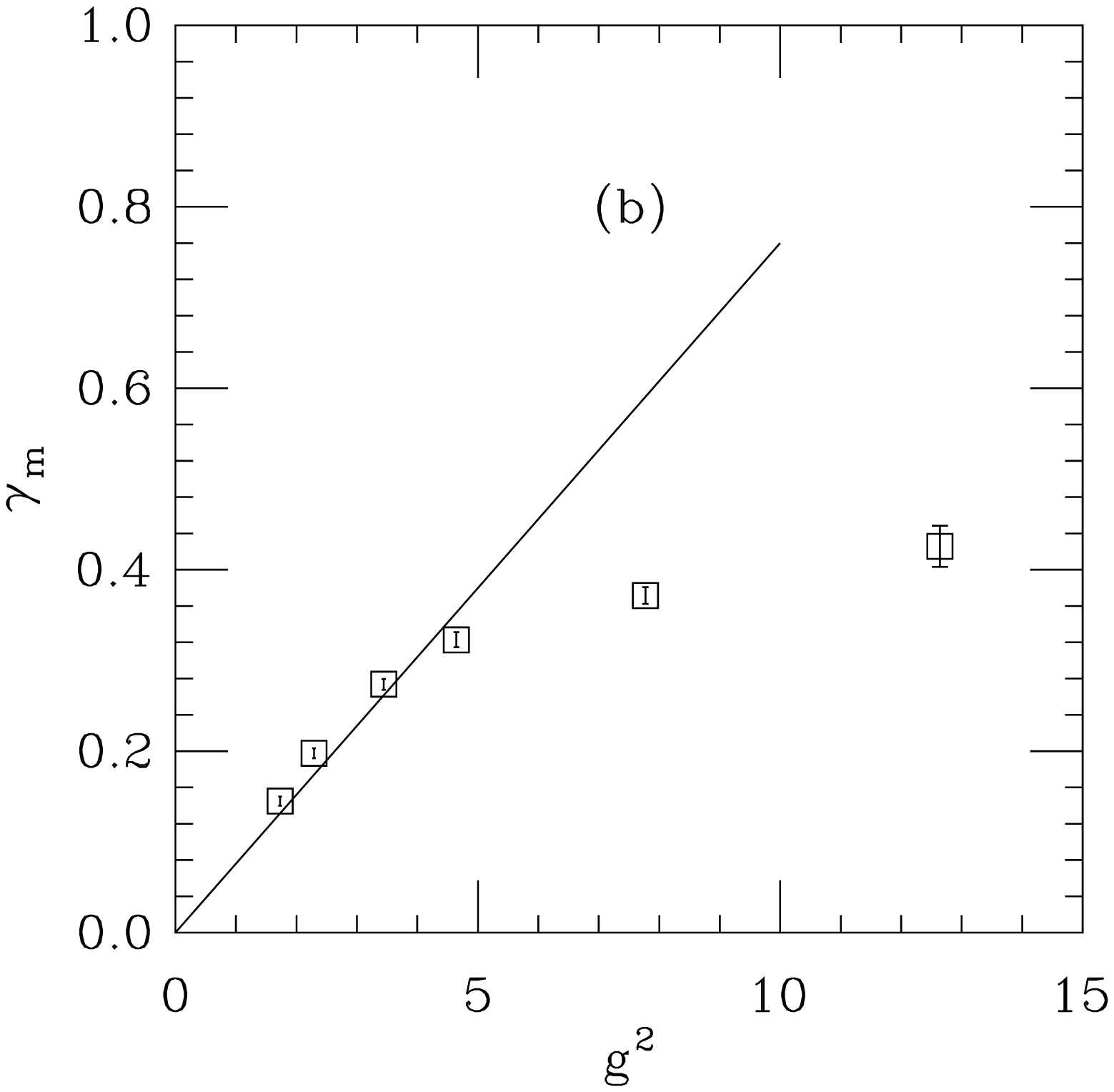}

\vspace*{5ex}

\includegraphics*[width=.48\textwidth]{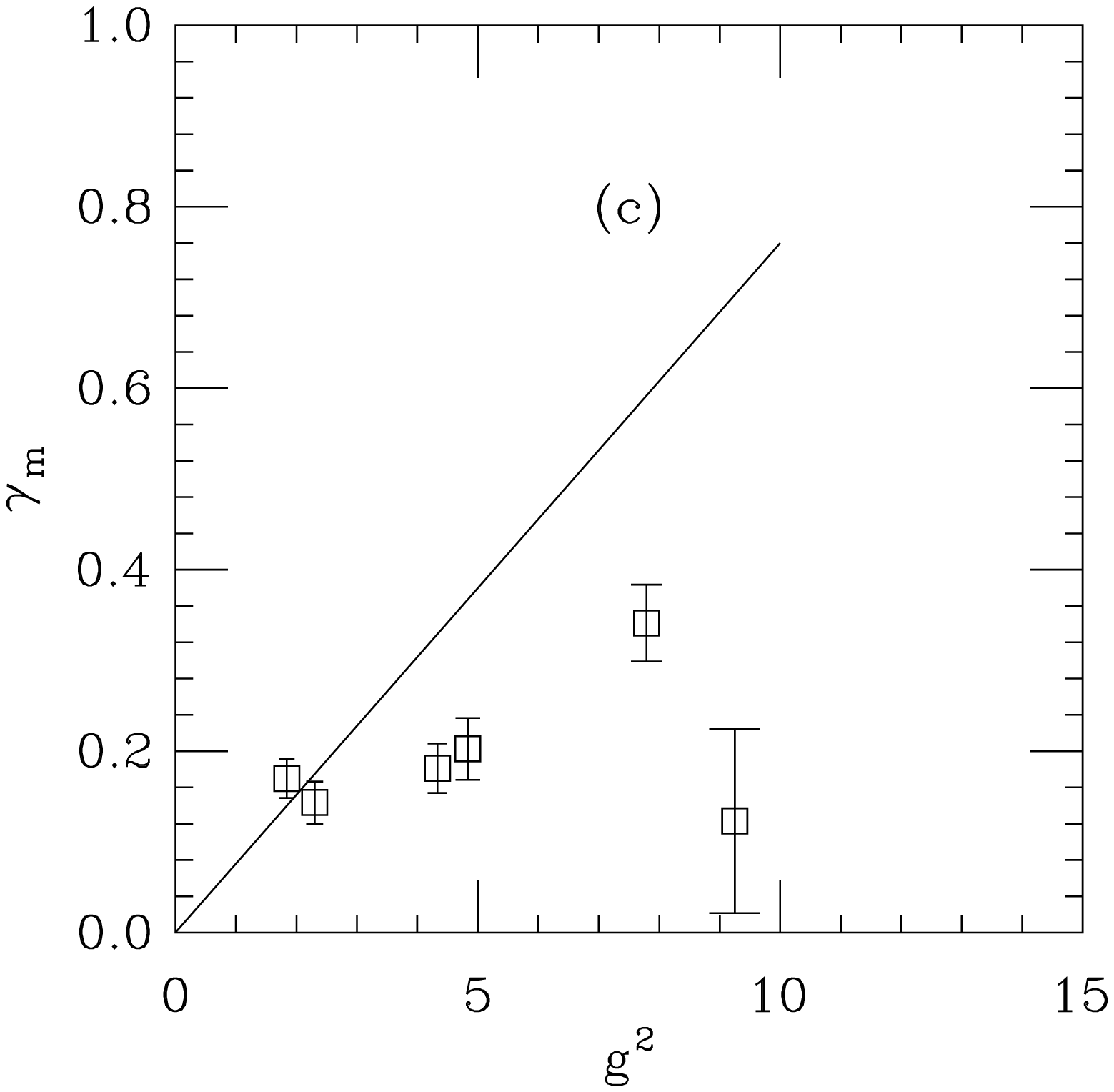}
\hspace{.02\textwidth}
\includegraphics*[width=.48\textwidth]{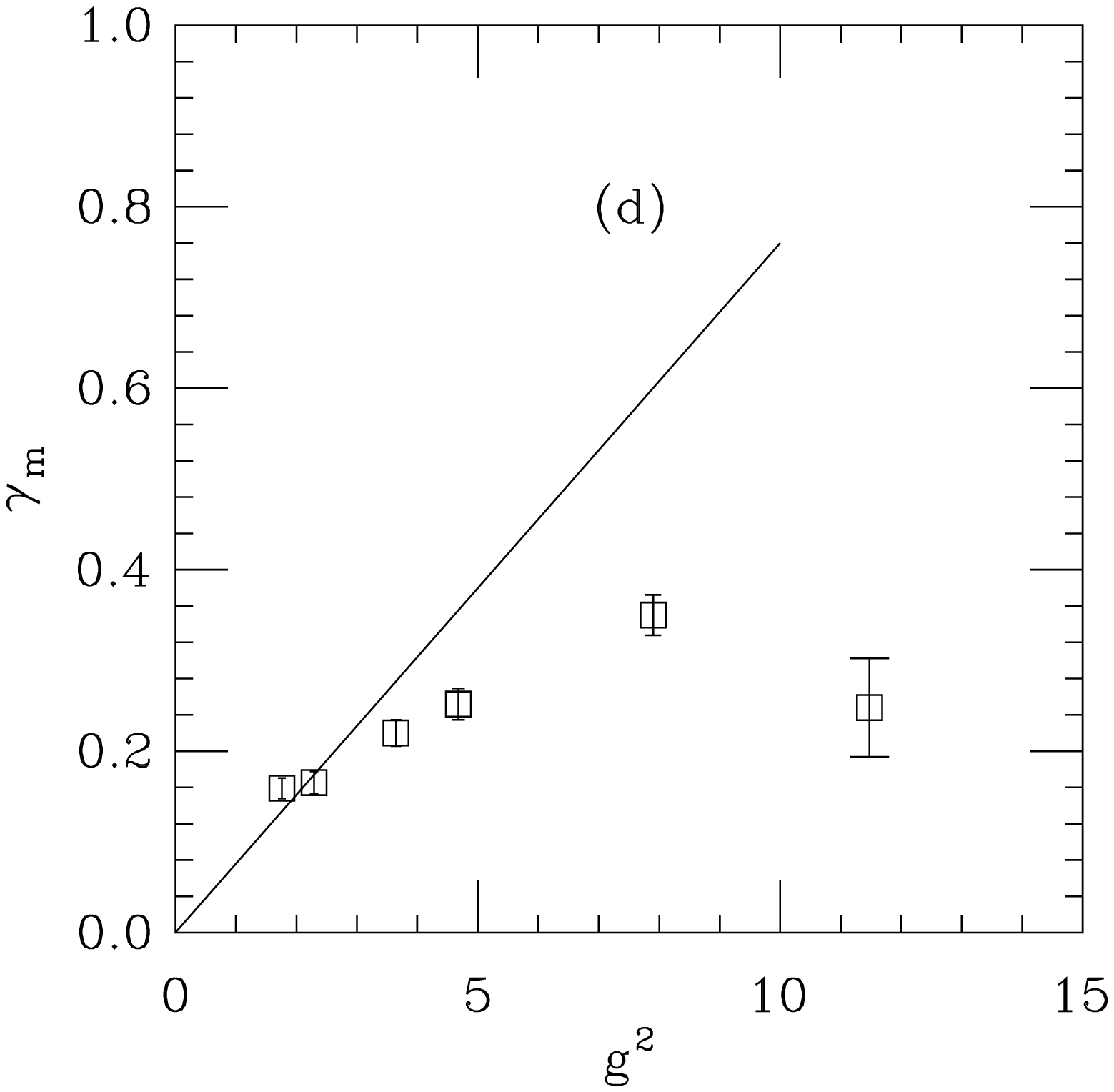}
\vspace*{0ex}
\end{center}
\caption{$\gamma_m$ from fits of $\log Z_p$ 
plotted as a function of $g^2(L=8a)$:
(a) Squares show fits to $a - \gamma_m\log x$, where $x=L/8a$,
using all four volumes;
circles, using $L/a=8$, 12, 16 only.
(b) Fits to $a - \gamma_m\log x + c(\log x)^2$.
(c) Fits to $a - \gamma_m\log x + c/x$.
(d) Fits to $a - \gamma_m\log x + c/x^2$.
The line in each case is the one-loop result.
\label{fig:effgammasu2m}}
\end{figure*}

As we did for the running coupling, we can fit $\log Z_P(L)$ to various
functional forms.
Again we begin with the simple linear behavior of \Eq{eq:lgamma}; we proceed
to add corrections with the aim of testing whether deviations reflect continuum corrections or lattice artifacts.  We use the same fit functions, given in
\Eq{grp:fits}, with the change that $u$ in these equations
now stands for $\log Z_P(L)$.  The reasoning that led to the identification
of the fit parameter $b$ with $R(g^2)$ in Sec.~\ref{2fit} now leads
to its identification with $-\gamma_m(g^2)$.
The results are shown in Fig.~\ref{fig:effgammasu2m}.
In all cases we plot $\gamma_m$ as a function of $g^2(L=8a)$, where the
latter was obtained from the parallel fit type in Sec.~\ref{2fit}.

%%%%%%%%%%%%%%%%%%%%%%%%%%%%%%%%%%%%%%%%%%%%%%%%%%%%%%%%%%%%%%%%%%%%%
%%%%%%%%%%%%%%%%%%%%%%%%%%%%%%%%%%%%%%%%%%%%%%%%%%%%%%%%%%%%%%%%%%%%%

Our data for $Z_p$
are far more precise than the data for $1/g^2$ given above.
This is reflected in the much smaller error bars in Fig.~\ref{fig:effgammasu2m}
compared to Fig.~\ref{fig:bfndatam}.  At the same time,
this leads to rather high values of $\chi^2$/dof (in the range 3--4) in the
linear fits~(\ref{eq:lgamma}), whether we use three or four volumes.
The $\chi^2$ is good for the other three fits,
again excepting $\beta=1.5$ which gives $\chi^2/\text{dof}\simeq6$.

\begin{figure}
\begin{center}
\includegraphics[width=\columnwidth,clip]{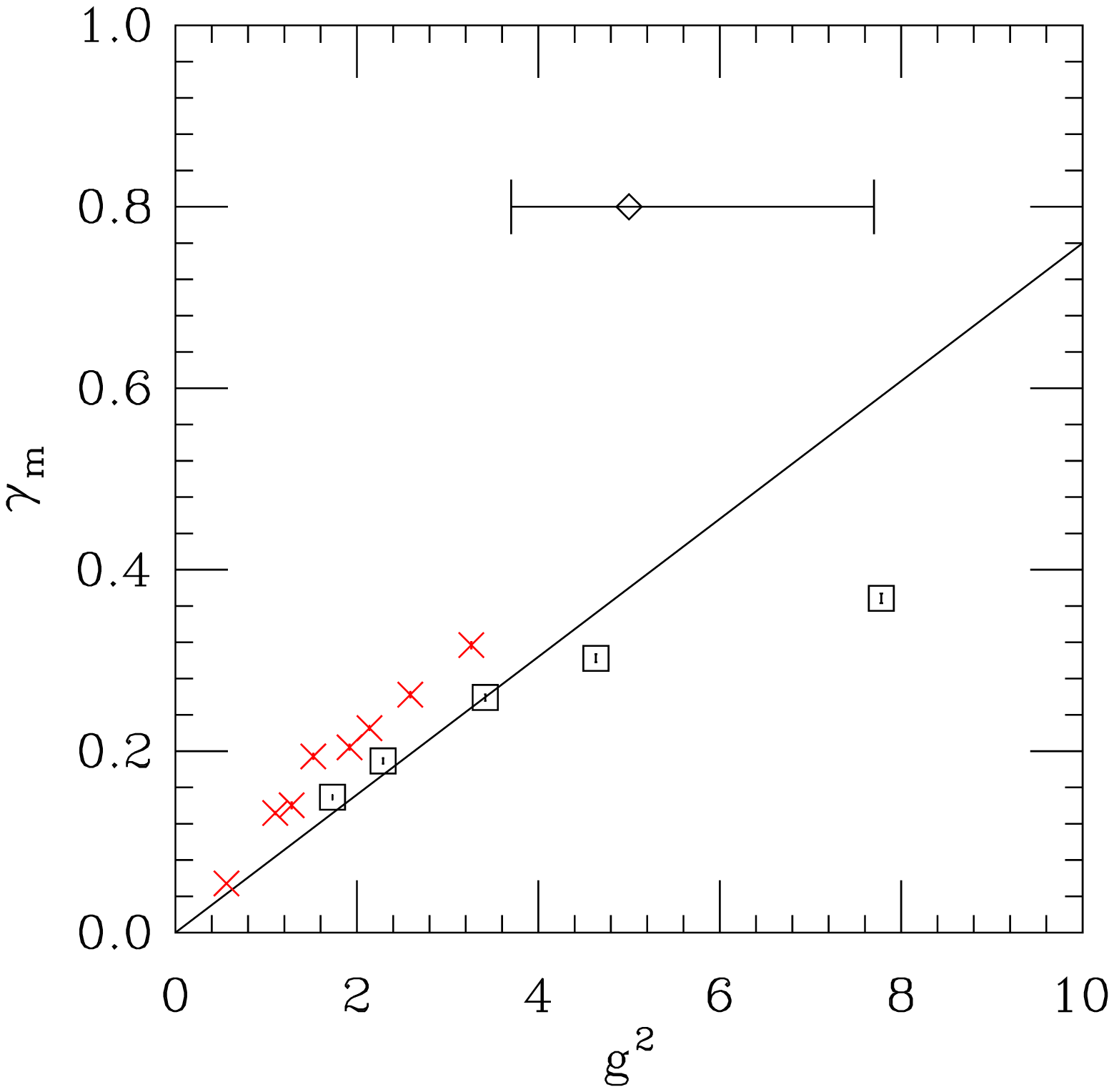}
\end{center}
\caption{Mass anomalous dimension $\gamma_m(g^2)$ from
the linear fit (\ref{eq:lgamma}),
which are the squares in Fig.~\ref{fig:effgammasu2m}(a).
The horizontal bar at the top marks our result~(\ref{gstar})
for $g_*^2$.
The crosses are the data of Bursa, et al.~\cite{Bursa:2009we},
analyzed with the same linear fit.
Again the diagonal line is the perturbative result.
\label{fig:gammamg2}}
\end{figure}

Finally we estimate $\gamma_m(g_*)$, the mass anomalous
dimension at the IRFP.
The uncertainty in $\gamma_m(g_*)$ must
reflect the uncertainties in the determination of $g_*$ itself.
In Fig.~\ref{fig:gammamg2} we have plotted again our results for the linear
fit to all four volumes.  The horizontal bar at the top of the figure
indicates our result for $g_*^2$, \Eq{gstar}.
We see that the right and left ends of the error bar
almost coincide with the points
corresponding to $\beta=2.0$ (on the weak-coupling side) and $\beta=1.5$
(on the strong-coupling side).  This observation renders unnecessary any interpolation of the curve. In view of the monotonicity of
$\gamma_m(g)$, we simply take the values of $\gamma_m$ at
these two couplings to mark off the uncertainty of
$\gamma_m(g_*)$, concluding that
\bee
\label{gammastar}
\gamma_m(g_*) = 0.31(6) .
\ee
Roughly $5/6$ of the error comes from the
uncertainty in $g_*$.  The small statistical and
systematic errors of $\gamma_m$ itself are responsible for the rest.
Based on the same reasoning as in Sec.~\ref{2fit},
we took the spread of values
obtained using the two linear fits [panel (a) of Fig.~\ref{fig:effgammasu2m}]
and fit~(\ref{fit:b}) [panel (b)] as a measure of the
systematic uncertainty in $\gamma_m$.

Fig.~\ref{fig:gammamg2} also shows a comparison of our data to those
of Bursa \textit{et al.}~\cite{Bursa:2009we}, where we have applied to their
data the linear fit~(\ref{eq:lgamma}). (Their published graphs include an additional
 large systematic uncertainty.) Our data lie slightly below theirs.

%%%%%%%%%%%%%%%%%%%%%%%%%%%%%%%%%%%%%%%%%%%%%%%%%%%%%%%%%%%%%%%%%%%%%
%%%%%%%%%%%%%%%%%%%%%%%%%%%%%%%%%%%%%%%%%%%%%%%%%%%%%%%%%%%%%%%%%%%%%

%%%%%%%%%%%%%%%%%%%%%%%%%%%%%%%%%%%%%%%%%%%%%%%%%%%%%%%%%%%%%%%%%%%%%
\section{Discussion \label{sec:last}}
%%%%%%%%%%%%%%%%%%%%%%%%%%%%%%%%%%%%%%%%%%%%%%%%%%%%%%%%%%%%%%%%%%%%%

Our simulations show that SU(2) gauge theory coupled to two flavors of adjoint fermions
lies inside the conformal window. Using  Schr\"odinger functional techniques,
we have determined its IRFP and measured two exponents, the mass anomalous dimension
$\gamma_m$ and the exponent $y_g$ of the (irrelevant) gauge coupling. Both are small.

Because the massless theory is conformal, not confining, it is not a candidate
for a technicolor theory. 
Even if the theory were to be deformed in a way that destroys the IRFP, the
small anomalous dimension would spell trouble for its application as a
technicolor theory.

We are aware of several estimates of $g_*$ and $\gamma_m(g_*)$ in the literature.
The two-loop zero of the beta function occurs at $g_*^2=7.9$ which is right on the edge of our
quoted range. Our central value is at  weaker coupling, $g_*^2=5.0$.

The location of the IRFP is scheme dependent, of course.
The previous lattice study with the best data,
which also used
the SF coupling to find a fixed point, is that of Hietanen {\em et al.}~\cite{Hietanen:2009az}.
Their published analysis combines data from all $L$'s and bare parameter values
into a single fitting function. They
quote fitted values of $g_*^2$ in the range 2.0--3.2, or $1/g_*^2$ in the range 0.3--0.5.
We are in mild disagreement with their results.
 Their lattice action uses unimproved Wilson fermions with no clover term,
and hence is susceptible to $O(a)$ discretization errors. The line of first order
 transitions also extends
farther into weaker coupling in their case; they have no data below  $1/g^2 =0.22$.

As seen in Fig.~\ref{fig:gammamg2}, our results for $\gamma_m(g^2)$ agree with the earlier determination of
Bursa {\em et al.}~\cite{Bursa:2009we}, where we overlap in couplings.
We also agree with the MCRG result of Catterall {\em et al.}~\cite{Catterall:2010du}, $\gamma_m(g_*)=0.49(13)$, given its larger uncertainty.
Del Debbio {\em et al.}~\cite{DelDebbio:2010hu,DelDebbio:2010hx} use the scaling of spectral observables with fermion mass to 
compute $\gamma_m$ at $g^2\simeq 3$, finding small values [0.05--0.20 and 0.22(6) respectively].
These results also lie on the lowest-order perturbative curve.

We find that
$\gamma_m(g^2)$ deviates from the lowest-order perturbative formula for $g^2\agt4$. 
As we saw for SU(3) with sextet fermions,
the numerical results lie below the curve.

Now let us consider analytic predictions for  $\gamma_m(g_*)$.
Ryttov and Shrock \cite{Ryttov:2010iz} have an extensive tabulation of
perturbative results up to four loops in the $\overline{MS}$ scheme
(see also~\cite{Pica:2010xq}).
Recall that only anomalous dimensions measured at fixed points are
scheme-independent.
Perturbative predictions of course depend on the order of perturbation theory.
They tabulate $\gamma_m(g_*) = 0.820$, 0.543, 0.500 for two, three, and four loops. 
They also give a 
prediction based on solving a Schwinger--Dyson equation, $\gamma_m(g_*) =0.653$.
The Ryttov--Sannino all-orders beta function~\cite{Ryttov:2007cx} gives $\gamma_m(g_*)=0.75$.
Pica and Sannino offer another all-orders beta function \cite{Pica:2010mt} which gives $\gamma_m(g_*)=0.46$.
All these numbers are too high to agree with our result. 

There is an extensive literature attempting to relate the location of the bottom of
the conformal window to a large value for $\gamma_m$
(see \cite{Ryttov:2007cx,Kaplan:2009kr} and references therein).
Our results indicate that
models using the SU(2) gauge group and adjoint fermions are not relevant
to that literature: the parameter space accessible to exploration is
too granular. Our $\gamma_m$ is small.
One might want to decrease $N_f$ in the hope that $\gamma_m$ would grow as one approaches
the bottom of the conformal window.
This is clearly not possible here:
The $N_f=1$ theory has $b_2>0$ and is probably confining.

Finally, we point out that
the use of an improved action, which smooth the gauge field fluctuations as seen by
the fermions,
made this project feasible. It shifts the location of the strong-coupling transition
deeper into strong coupling than could be achieved with the simple Wilson action,
allowing us access to the strong coupling side of the IRFP.

%%%%%%%%%%%%%%%%%%%%%%%%%%%%%%%%%%%%%%%%%%%%%%%%%%%%%%%%%%%%%%%%%%%%%
\begin{acknowledgments}
%%%%%%%%%%%%%%%%%%%%%%%%%%%%%%%%%%%%%%%%%%%%%%%%%%%%%%%%%%%%%%%%%%%%%%
B.~S. and Y.~S. thank the University of Colorado for hospitality.
This work was supported in part by the Israel Science Foundation
under grant no.~423/09 and by the U.~S. Department of Energy.
Computations were done on clusters at the University of Colorado and Tel Aviv University.
Additional computations were done on facilities of the USQCD Collaboration at Fermilab,
which are funded by the Office of Science of the U.~S. Department of Energy.
Our computer code is based on the publicly available package of the
 MILC collaboration~\cite{MILC}.
The code for hypercubic smearing was adapted from a program written by A.~Hasenfratz,
R.~Hoffmann and S.~Schaefer~\cite{Hasenfratz:2008ce}.
%%%%%%%%%%%%%%%%%%%%%%%%%%%%%%%%%%%%%%%%%%%%%%%%%%%%%%%%%%%%%%%%%%%%%
\end{acknowledgments}
%%%%%%%%%%%%%%%%%%%%%%%%%%%%%%%%%%%%%%%%%%%%%%%%%%%%%%%%%%%%%%%%%%%%%

%%%%%%%%%%%%%%%%%%%%%%%%%%%%%%%%%%%%%%%%%%%%%%%%%%%%%%%%%%%%%%%%%%%%%
\appendix*
\section{Determining $\kappa_c$}
%%%%%%%%%%%%%%%%%%%%%%%%%%%%%%%%%%%%%%%%%%%%%%%%%%%%%%%%%%%%%%%%%%%%%

For each value of the bare coupling $\beta$ we determined $\kappa_c$ by demanding $m_q=0$ for $L=12a$.
We calculated $m_q$ in a series of short runs over a range of $\kappa$'s, eventually confirming $m_q=0$ in a run of several hundred trajectories.
When we used this value of $\kappa_c$ in the much longer runs that yielded the SF coupling $g^2$, the error bar on $m_q$ naturally shrank and thus the final result for $m_q$ is always a little bit off zero.

\begin{figure}
\begin{center}
\includegraphics[width=\columnwidth,clip]{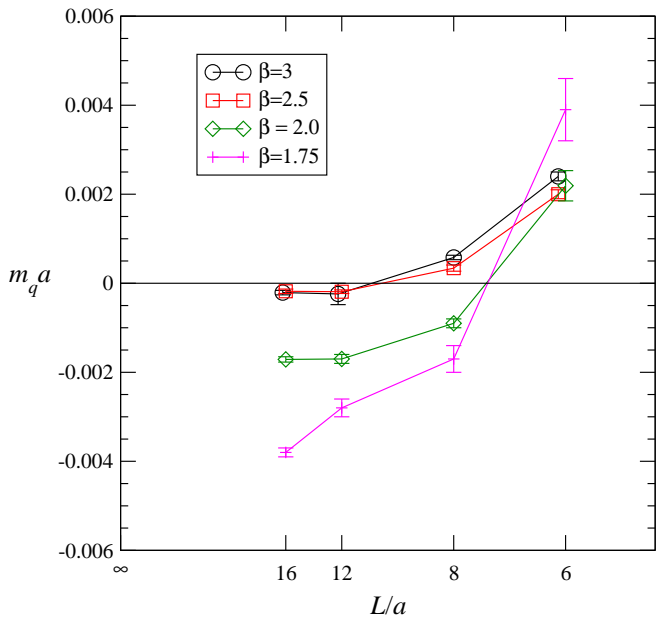}
\end{center}
\caption{Volume dependence of the AWI mass $m_q$ at fixed $\beta$, where $\kappa$ was fixed by a preliminary determination of $m_q=0$ on the $12^4$ lattice.
\label{fig:mq1}}
\end{figure}

We used the same values of $(\beta,\kappa_c)$ for different lattice volumes, so as to keep the lattice spacing fixed for the SF calculation.
In the weak-coupling region, the volume dependence of $\kappa_c$ is weak, as seen in Fig.~\ref{fig:mq1}.
%%%%%%%%%%%%%%%%%%%%%%%%%%%%%%%%%%%%%%%%%%%%%%%%%%%%%%%%%%%%%%%%%%%%%
%%%%%%%%%%%%%%%%%%%%%%%%%%%%%%%%%%%%%%%%%%%%%%%%%%%%%%%%%%%%%%%%%%%%%
%%%%%%%%%%%%%%%%%%%%%%%%%%%%%%%%%%%%%%%%%%%%%%%%%%%%%%%%%%%%%%%%%%%%%
\begin{figure}
\begin{center}
\includegraphics[width=\columnwidth,clip]{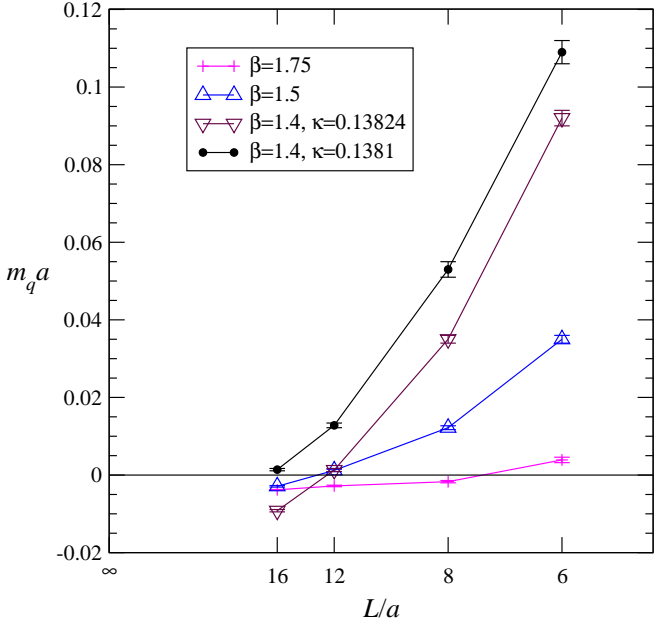}
\end{center}
\caption{Same as Fig.~\ref{fig:mq1}, but at stronger bare couplings $\beta$. The data for $\beta=1.75$ are the same as in Fig.~\ref{fig:mq1}, to show the change in vertical scale.
\label{fig:mq2}}
\end{figure}
%%%%%%%%%%%%%%%%%%%%%%%%%%%%%%%%%%%%%%%%%%%%%%%%%%%%%%%%%%%%%%%%%%%%%
In particular, $m_q$ on the $16^4$ lattice is small enough that we can say that $m_q\simeq0$ in the infinite-volume limit.
This is not the case, however, at stronger couplings (Fig.~\ref{fig:mq2}).
Of particular concern is the large value of $m_q$ for the strongest coupling, $\beta=1.4$, on the largest lattice, $L=16a$; the value $\kappa_c=0.13284$ is perfectly adequate for $L=12a$ but not for $L=16a$ (see Table~\ref{table:mq1}).
%%%%%%%%%%%%%%%%%%%%%%%%%%%%%%%%%%%%%%%%%%%%%%%%%%%%%%%%%%%%%%%%%%%%%
\begin{table*}
\caption{AWI masses $m_qa$ at $\beta=1.4$ for the two $\kappa$ values considered.}
\begin{center}
\begin{ruledtabular}
\begin{tabular}{ddrrrr}
\beta & \kappa &\multicolumn{4}{c}{$m_qa$}\\
\cline{3-6}
&&              $L=6a$   & $L=8a$   & $L=12a$   & $L=16a$    \\
\hline
1.4 & 0.13824 (\equiv\kappa_c)
              & 0.092(2) & 0.035(1) & 0.0013(4) & -0.0092(3) \\
1.4 & 0.1381  & 0.109(3) & 0.053(2) & 0.0128(6) &  0.0014(3) \\
\end{tabular}
\end{ruledtabular}
\end{center}
\label{table:mq1}
\end{table*}
%%%%%%%%%%%%%%%%%%%%%%%%%%%%%%%%%%%%%%%%%%%%%%%%%%%%%%%%%%%%%%%%%%%%%
%%%%%%%%%%%%%%%%%%%%%%%%%%%%%%%%%%%%%%%%%%%%%%%%%%%%%%%%%%%%%%%%%%%%%
\begin{table*}
\caption{Schr\"odinger functional couplings $1/g^2$ at $\beta=1.4$ for the two $\kappa$ values.}
\begin{center}
\begin{ruledtabular}
\begin{tabular}{ddrrrr}
\beta & \kappa &\multicolumn{4}{c}{$1/g^2$}\\
\cline{3-6}
&&              $L=6a$    & $L=8a$     & $L=12a$    & $L=16a$    \\
\hline
1.4 & 0.13824 (\equiv\kappa_c)
             & 0.0655(29) & 0.0790(21) & 0.0950(27) & 0.1035(34) \\
1.4 & 0.1381 & 0.0610(21) & 0.0722(24) & 0.0891(34) & 0.0926(34) \\
\end{tabular}
\end{ruledtabular}
\end{center}
\label{table:mq2}
\end{table*}
%%%%%%%%%%%%%%%%%%%%%%%%%%%%%%%%%%%%%%%%%%%%%%%%%%%%%%%%%%%%%%%%%%%%%

In principle, one could fix $\kappa_c$ by demanding that $m_q\to0$ in the infinite volume limit, which can be done by fitting to finite-volume results at various values of $\kappa$.
As is clear from Fig.~\ref{fig:mq2}, the finite-lattice corrections to $m_q$ are proportional to $(a/L)^2$, as may be expected for the clover action.
This procedure would involve lengthy calculations, however, since the error bars shown in the figure are only attainable with the statistics of a full SF simulation.
We decided instead to check on the sensitivity of our results to a shift in $\kappa$ of the order that might be required by this procedure.

Table~\ref{table:mq1} and Fig.~\ref{fig:mq2} show the values of $m_qa$ calculated at the nominal $\kappa_c=0.13824$ and at a shifted value $\kappa=0.1381$ that gives a small value of $m_qa$ for the largest lattice $L=16a$.
Table~\ref{table:mq2} lists the SF couplings $1/g^2$ for all four volumes at both values of $\kappa$.
While the shift in $\kappa$ does induce a significant and systematic shift in $1/g^2$, the change in the DBF is much smaller than the statistical error, as is seen in Fig.~\ref{fig:SU2DBF2}.
A similar result obtains for $Z_P$ (Table~\ref{table:mq3});
the ratios $\sigma_P$ are unaffected and thus the estimate
of $\gamma_m$ is unaffected as well.
We conclude that our results are insensitive to such shifts in $\kappa$, even at the strongest coupling studied.
Note that we have not made use of any of the data at $\beta=1.4$ in determining
$g_*$ or $\gamma_m(g_*)$.
%%%%%%%%%%%%%%%%%%%%%%%%%%%%%%%%%%%%%%%%%%%%%%%%%%%%%%%%%%%%%%%%%%%%%
\begin{table*}
\caption{Pseudoscalar renormalization constant $Z_P$ at $\beta=1.4$ for the two $\kappa$ values.}
\begin{center}
\begin{ruledtabular}
\begin{tabular}{ddrrrr}
\beta & \kappa &\multicolumn{4}{c}{$Z_P$}\\
\cline{3-6}
&&              $L=6a$   & $L=8a$          & $L=12a$  & $L=16a$         \\
\hline
1.4 & 0.13824 (\equiv\kappa_c)
              & 0.250(3) & 0.219(2)\ \ \ \ & 0.188(1) & 0.173(2)\ \ \ \ \\
1.4 & 0.1381  & 0.237(2) & 0.2105(13)      & 0.182(3) & 0.1645(15)      \\
\end{tabular}
\end{ruledtabular}
\end{center}
\label{table:mq3}
\end{table*}
%%%%%%%%%%%%%%%%%%%%%%%%%%%%%%%%%%%%%%%%%%%%%%%%%%%%%%%%%%%%%%%%%%%%%

%%%%%%%%%%%%%%%%%%%%%%%%%%%%%%%%%%%%%%%%%%%%%%%%%%%%%%%%%%%%%%%%%%%%%

%%%%%%%%%%%%%%%%%%%%%%%%%%%%%%%%%%%%%%%%%%%%%%%%%%%%%%%%%%%%%%%%%%%%%
\end{document}